\documentclass[preprint,a4paper,12pt,3p]{elsarticle}
\biboptions{square,sort&compress} 
\usepackage{scalefnt}
\usepackage{hyperref}
\usepackage{todonotes}
\usepackage{graphicx}
\usepackage{graphics}
\usepackage[mathscr]{eucal}
\usepackage{amssymb}
\usepackage{amsmath}
\usepackage{tabls} 
\usepackage{multirow}
\usepackage{wasysym}
\usepackage{float}
\usepackage{listings}
\usepackage{tikz}
\tikzstyle{every picture}+=[remember picture]
\tikzstyle{na} = [baseline=-.5ex]
\usetikzlibrary{arrows,shapes,backgrounds}
\tikzstyle{background grid}=[draw, black!50,step=.5cm]

\newcommand{\fs}[1]{#1{\abbrev FS}}
\newcommand{\mass}[1]{m_{#1}}

\newcommand{\lhc}{{\abbrev LHC}}
\newcommand{\sm}{{\abbrev SM}}
\newcommand{\mssm}{{\abbrev MSSM}}
\newcommand{\susy}{{\abbrev SUSY}}
\newcommand{\lo}{{\abbrev LO}}
\newcommand{\nlo}{{\abbrev NLO}}
\newcommand{\nblo}{{\abbrev (N)LO}}
\newcommand{\nnlo}{{\abbrev NNLO}}
\newcommand{\nbnlo}{{\abbrev (N)NLO}}
\newcommand{\nbnblo}{{\abbrev (NN)LO}}
\newcommand{\qcd}{{\abbrev QCD}}
\newcommand{\msbar}{\overline{\text{MS}}}
\newcommand{\drbar}{\overline{\text{DR}}}
\newcommand{\OS}{\text{OS}}
\newcommand{\EW}{\text{EW}}
\newcommand{\abbrev}{\rm\scalefont{.9}}
\newcommand{\muF}{\mu_{\rm F}}
\newcommand{\muR}{\mu_{\rm R}}
\newcommand{\muD}{\mu_{\rm d}}

\newcommand{\mhiggs}{\mass{H}}
\newcommand{\mphi}{\mass{\phi}}
\newcommand{\mbottom}{\mass{b}}
\newcommand{\mcharm}{\mass{c}}
\newcommand{\mtop}{\mass{t}}
\newcommand{\mstop}[1]{\mass{\tilde{t}#1}}
\newcommand{\msbottom}[1]{\mass{\tilde{b}#1}}
\newcommand{\msquark}[1]{\mass{\tilde{q}#1}}
\newcommand{\mgluino}{\mass{\tilde{g}}}
\newcommand{\pdf}{{\abbrev PDF}}

\newcommand{\eqn}[1]{Eq.\,(\ref{#1})}
\newcommand{\neqn}[1]{Eqs.\,(\ref{#1})}
\newcommand{\fig}[1]{Fig.\,\ref{#1}}
\newcommand{\tab}[1]{Tab.\,\ref{#1}}
\newcommand{\sct}[1]{Section~\ref{#1}}
\newcommand{\scts}[1]{Sections~\ref{#1}}

\newcommand{\order}[1]{{\cal O}(#1)}

\newcommand{\citere}[1]{Ref.\,\cite{#1}}
\newcommand{\citeres}[1]{Refs.\,\cite{#1}}
\renewcommand{\Re}{{\rm Re}}

\newcommand{\sushi}{{\tt SusHi}}
\newcommand{\tq}{{\tilde{q}}}
\newcommand{\ttop}{{\tilde{t}}}
\newcommand{\tbot}{{\tilde{b}}}

\newcommand{\red}[1]{\textcolor{red}{#1}}

\def\Tiny{ \font\Tinyfont = cmr10 at 3.5pt \relax  \Tinyfont}
\newcommand{\nobar}[1]{%
{\kern-0.5em\vbox to+8.5pt{\hbox{\textbf{\Tiny(}}}\kern-0.1em \overline{#1}} \kern-0.3em\vbox to+8.5pt{\hbox{\textbf{\Tiny)}}}}

\tikzstyle{boxblue} = [draw=blue, fill=blue!10, very thick,
    rectangle, rounded corners, inner sep=5pt, inner ysep=5pt]

\tikzstyle{boxred} = [draw=red, fill=red!10, very thick,
    rectangle, rounded corners, inner sep=5pt, inner ysep=5pt]

\tikzstyle{boxgreen} = [draw=green, fill=green!10, very thick,
    rectangle, rounded corners, inner sep=5pt, inner ysep=5pt]

\begin{document}

\begin{frontmatter}

\title{
\vspace*{-6em}
  \begin{flushright}
    {\sf\small March 2013 --- WUB/12-28, LPN12-134}
  \end{flushright}
\vspace*{2em}
\sushi: A program for the calculation of Higgs production in gluon fusion and
bottom-quark annihilation in the Standard Model and the MSSM}

\author{Robert V. Harlander}
\ead{harlander@physik.uni-wuppertal.de}

\author{Stefan Liebler}
\ead{sliebler@physik.uni-wuppertal.de}

\author{Hendrik Mantler}
\ead{hendrik.mantler@uni-wuppertal.de}

\address{
Fachbereich C, Bergische Universit\"at Wuppertal\\
42097 Wuppertal, Germany
}


\begin{abstract}
This article describes the code \sushi{} (for
``\underline{Su}per\underline{s}ymmetric \underline{Hi}ggs'')
\cite{sushiaddress} which calculates the cross sections
$pp/p\overline{p}\rightarrow\phi+X$ in gluon fusion and bottom-quark
annihilation in the \sm{} and the \mssm{}, where $\phi$ is any of the
neutral Higgs bosons within these models. Apart from inclusive cross
sections up to \nnlo{} \qcd{}, differential cross sections with respect
to the Higgs transverse momentum $p_T$ and (pseudo-)rapidity $y(\eta)$
can be calculated through \nlo{} \qcd{}. In the case of gluon fusion,
\sushi{} contains \nlo{} \qcd{} contributions from the third family of
quarks and squarks, \nnlo{} corrections due to top-quarks, approximate
\nnlo{} corrections due to top-squarks, and electro-weak effects. It
supports various renormalization schemes for the sbottom sector and the
bottom Yukawa coupling, as well as resummation effects of higher order
$\tan\beta$-enhanced sbottom contributions.  \sushi{} provides a link to
{\tt FeynHiggs} for the calculation of the Higgs masses.
\end{abstract}

\end{frontmatter}

\tableofcontents



\section{Introduction}
\label{sec:introduction}

The recent observation of a new boson at the Large Hadron Collider (\lhc{})
\cite{Aad:2012gk,Chatrchyan:2012gu} has opened a new chapter for Higgs
phenomenology~\cite{Dittmaier:2012nh}.  For the clear identification of
this particle, precise predictions for Higgs production and decay will
be absolutely essential. The current status of these efforts is
collected in the reports of the \lhc{} Higgs cross section working group
\cite{Dittmaier:2011ti,Dittmaier:2012vm}.

The main production mechanism for a Standard Model (\sm{}) Higgs boson
at a hadron collider is gluon fusion, where the gluon-Higgs coupling is
mediated mostly by virtual top- and bottom-quarks ($\sim 7\%$).  The
total inclusive cross section is known through next-to-next-to-leading
order (\nnlo{}) in quantum chromodynamics
(\qcd{})~\citep{Georgi:1977gs,Djouadi:1991tka,
  Dawson:1990zj,Spira:1995rr,Harlander:2002wh,Anastasiou:2002yz,
  Ravindran:2003um,Marzani:2008az,Harlander:2009mq,Pak:2009dg,
  Harlander:2009my,Pak:2011hs}.  Even higher order \qcd{} effects have
been calculated~\cite{Catani:2003zt,
  Moch:2005ky,Idilbi:2005ni,Idilbi:2006dg,Ravindran:2006cg,Ahrens:2008nc}
through resummation; electro-weak contributions reach up to $8\%$ with
respect to the leading order (\lo{}) cross
section~\cite{Actis:2008ug,Aglietti:2004nj,Bonciani:2010ms}.

The gluon-fusion mechanism for the neutral Higgs bosons in the Minimal
Supersymmetric Standard Model (\mssm{}) is mediated by quarks and their
superpartners, the squarks. For the {\abbrev CP}-even \mssm{} Higgs
bosons $h,H$, the \qcd{} effects due to quarks can be simply taken over
from the \sm{} by a rescaling of the cross section with the
corresponding modified Yukawa couplings. For the {\abbrev CP}-odd Higgs
boson $A$, the \nnlo{} \qcd{} corrections to the quark-induced total
inclusive cross section have been calculated in
Refs.\,\cite{Harlander:2002vv,Anastasiou:2002wq,Pak:2011hs}.

Squark contributions to the gluon-Higgs coupling are typically
suppressed by powers of $\mass{q}/\mass{\tilde{q}}$, and thus of
importance mostly for small to moderate squark masses
$\mass{\tilde{q}}$.  For the {\abbrev CP}-odd Higgs boson $A$, they are
absent at \lo{}.\footnote{Comprehensive reviews of the Higgs theory
within the \sm{} and the \mssm{} can be found in
Refs.\,\cite{Djouadi:2005gi,Djouadi:2005gj}.}

The individual components for a calculation of \nlo{} \qcd{} corrections
in the \mssm{} are known: The \nlo{} cross section due to (s)top
(s)quarks, gluons, and gluinos was calculated for a {\abbrev CP}-even
Higgs boson with mass $\mphi$ in
\citeres{Harlander:2003bb,Harlander:2004tp,Degrassi:2008zj} by applying
an effective-theory approach in the limit
$\mphi\ll\mtop{},\mstop{},\mgluino$, similar to what is used in the
\sm{} at higher orders.  In this approach, even \nnlo{} effects have
been first approximated\,\cite{Harlander:2003kf}, and recently fully
calculated~\cite{Pak:2010cu,Pak:2012xr}. The analogous \nlo{} result for
a {\abbrev CP}-odd Higgs was first obtained in
\citere{Harlander:2005if}.  Due to the smallness of the bottom-quark
mass, these results cannot be transferred to the bottom/sbottom
sector. However, the limit $\mphi{},\mbottom\ll \msbottom{},\mgluino$ is
applicable in a large region of the parameter space, and the
corresponding results were presented in
Refs.\,\cite{Degrassi:2010eu,Harlander:2010wr}. Recently, more general
results for the cross sections, allowing for larger Higgs masses, were
obtained for the {\abbrev CP}-even and -odd Higgs bosons and both for
the top/stop and the bottom/sbottom sector, in
Refs.\,\cite{Degrassi:2011vq,Degrassi:2012vt}.  A fully numerical
calculation of the gluon-Higgs form factor for general
quark/squark/gluino/Higgs-masses has been reported in
Ref.\,\cite{Anastasiou:2008rm}. For the pure squark contributions, the
full Higgs-mass dependence of the \nlo{} contribution to the cross
section was presented (numerically and/or analytically) in
\citeres{Anastasiou:2006hc,Aglietti:2006tp,Muhlleitner:2006wx}.

For large values of the \mssm{} parameter $\tan\beta$, the coupling of
the light {\abbrev CP}-even Higgs boson to bottom-quarks is
significantly enhanced relative to the \sm{} Yukawa coupling, so
the bottom sector may be much more important for gluon fusion in the
\mssm{}.  In addition, an enhanced bottom-Higgs coupling increases the
cross section of another Higgs production mechanism in the \mssm{},
namely associated production with bottom-quarks,
$pp/p\overline{p}\rightarrow b\overline{b}\phi$. If the final state
quarks are not tagged, a suitable theoretical approach to the cross
section is the process $b\overline{b}\rightarrow \phi$, called
bottom-quark annihilation in what follows.\footnote{See
  Ref.\,\cite{bbh-santander} for a more detailed discussion.} It resums
terms of the form $\ln\mbottom/\mphi$ by means of $b$-parton
distribution functions (\pdf{}s) and was calculated up to \nnlo{} \qcd{}
in the \sm{}~\cite{Maltoni:2003pn,Harlander:2003ai}. The result can be
directly translated into the \mssm{} by rescaling it with the proper
bottom-Yukawa coupling; even the dominant sbottom effects can be taken
into account by an effective
coupling~\cite{Dittmaier:2006cz,Dawson:2011pe}.  \sushi{} evaluates both
the cross section for gluon fusion and bottom-quark annihilation.

For gluon fusion, \sushi{} includes results for all \nlo{} \qcd{}
contributions due to the third generation of quarks and squarks.  The
real corrections at \nlo{} are well-known; \sushi{} implements them
using the routines of Ref.\,\cite{Harlander:2010wr}.  For the virtual
corrections to the pure quark diagrams, it uses the analytic expression
of \citere{Harlander:2005rq} which was obtained from the integral
representation in \citere{Spira:1995rr}. Concerning the genuine virtual
supersymmetric (\susy{}) corrections, it employs the results of
Refs.\,\cite{Harlander:2003bb,Harlander:2004tp,Degrassi:2011vq,Degrassi:2012vt}
and \citeres{Degrassi:2010eu,Degrassi:2011vq} for the (s)top- and the
(s)bottom-mediated gluon-Higgs coupling, respectively. \nnlo{} \qcd{}
effects are taken into account for the top-quark induced gluon-Higgs
coupling~\cite{Harlander:2002wh,Anastasiou:2002yz,Ravindran:2003um,
  Harlander:2002vv,Anastasiou:2002wq}, and approximately for the
top/stop/gluino-induced one\,\cite{Harlander:2003kf} by using
{\tt ggh@nnlo}~\cite{ggh@nnlo}. Electro-weak
corrections\,\cite{Actis:2008ug,Aglietti:2004nj,Bonciani:2010ms} are
included as tabulated correction factors.
The cross section is provided in various renormalization schemes (in
particular in the sbottom sector), allowing for an on-shell, $\drbar$,
or a dependent renormalization of the soft-breaking parameter~$A_b$, for
example. In addition, the bottom-Yukawa coupling can be chosen on-shell
or in the $\msbar$-scheme. Higher-order sbottom effects can be
included through the parameter
$\Delta_b$~\cite{Banks:1987iu,Hall:1993gn,Hempfling:1993kv,Carena:1994bv,
Carena:1999py,Carena:2000uj}.

For the calculation of the bottom-quark annihilation cross section, \sushi{}
makes use of {\tt bbh@nnlo}~\cite{bbh@nnlo} and re-weights its results
by the \mssm{} couplings. It uses the {\tt LHAPDF}
library~\cite{Whalley:2005nh} which allows to conveniently switch
between different \pdf{} sets, and it can be linked to {\tt
  FeynHiggs}~\cite{Heinemeyer:1998np,Heinemeyer:1998yj,
  Degrassi:2002fi,Frank:2006yh} for the two-loop calculation of the
Higgs boson masses in the \mssm{}.

Apart from inclusive cross sections for gluon fusion and bottom-quark
annihilation, \sushi{} allows for (upper and lower) cuts on the
transverse momentum and/or the (pseudo-)rapidity of the outgoing scalar
$\phi$.  In case of gluon fusion, differential distributions with
respect to these kinematic variables can be obtained (for $p_T$ not too
small).  

Note that a number of codes for the calculation of Higgs cross sections
in the \sm{} and the \mssm{} exist, see
\citeres{Spira:1995mt,Bagnaschi:2011tu,
  Anastasiou:2011pi,Anastasiou:2009kn,Catani:2007vq,Catani:2008me,
  Anastasiou:2004xq}, for example. They overlap with \sushi{} to a
greater or lesser extent; the distinctive feature of \sushi{} is to
provide full \nlo{} \qcd{} (and partial \nnlo{} and electro-weak)
corrections for the dominant production mechanisms of the three neutral
Higgs bosons of the \mssm{}, both inclusive and differential, in various
renormalization schemes. Further details will be given below.

The remainder of this paper is organized as follows: In
\sct{sec:physics}, we present the physical background of \sushi{},
recalling the framework of the Higgs and quark/squark sectors in the
\mssm{} with special emphasis on the renormalization of the squark
sectors and the resummation of $\tan\beta$-enhanced sbottom corrections
in the bottom Yukawa coupling. Subsequently, we discuss the various
contributions for the calculation of the gluon-fusion and the
bottom-quark annihilation cross section as they enter in \sushi{}. We
briefly describe the kinematic variables for which cuts can be applied
and distributions be obtained.  In \sct{sec:program}, we describe the
program \sushi{}, in particular its workflow, installation, and usage,
as well as the input and output files.  Our conclusions are given in
\sct{sec:conclusions}. \ref{sec:higgssquark} contains the couplings of
the squarks to the Higgs bosons $\phi$.


\section{Physics background}
\label{sec:physics}


This section first introduces our notation for the relevant parts of the
\sm{} and the \mssm{}.  It describes the renormalization of the squark
sector and the possible choices for the bottom Yukawa coupling provided
in \sushi{}.


\subsection{Standard Model}

The \sm{} contains one scalar weak isospin doublet which corresponds to
a single physical particle $H$ (electric charge and spin zero). The pure
Higgs sector of the Lagrangian is determined by two parameters: the
vacuum expectation value $v\approx 246$\,GeV, and the mass of the Higgs
boson $\mhiggs$. The Yukawa couplings of the fermions to the Higgs boson
are given by $Y_f = \sqrt{2}\mass{f}/v$, where $\mass{f}$ is the fermion
mass. \sushi{} requires all input masses in the on-shell scheme, except
for the charm- and bottom-quark mass which has to be given as the
$\msbar$ mass $m_q(m_q)\equiv m_q^{\msbar}(m_q^{\msbar})$,
$q\in\{c,b\}$. The calculation of various internal bottom masses is
addressed in \sct{sec:bottommasses}.


\subsection{Supersymmetry}


The \mssm{} contains two Higgs doublets, named $H_d$ and $H_u$, which
develop the vacuum expectation values $v_d = v\cos\beta$ and
$v_u=v\sin\beta$, where the parameter $\beta$ is undetermined. They form
two {\abbrev CP}-even Higgs fields $h,H$, one {\abbrev CP}-odd (or
``pseudo-scalar'') Higgs field $A$, and two charged Higgs fields
$H^\pm$.  At lowest order, the mass spectrum of the Higgs sector is
determined by \sm{} parameters, $\tan\beta=v_u/v_d$, and the {\abbrev
  CP}-odd Higgs mass $\mass{A}$. Radiative corrections to the Higgs mass
spectrum are generally quite
large \cite{Ellis:1990nz,Okada:1990vk,Haber:1990aw}; currently, they
are known through three-loop
order \cite{Harlander:2008ju,Kant:2010tf,Martin:2007pg}. \sushi{}
provides a way to conveniently take into account two-loop corrections by
linking to the program
{\tt FeynHiggs} \cite{Heinemeyer:1998np,Heinemeyer:1998yj,
  Degrassi:2002fi,Frank:2006yh}. This is an optional feature, however;
any Higgs mass can be given as an input to \sushi{}.

The mixing of the isospin to the Higgs mass eigenstates in the {\abbrev
  CP}-even sector is governed by the angle $\alpha$. Together with
$\beta$, it determines the relative strengths of the Higgs boson
couplings $g_f^\phi$ ($\phi\in\lbrace h,H,A\rbrace$) to the \sm{}
fermions with respect to the \sm{} Higgs boson couplings (see
\ref{sec:higgssquark}), which thus enter the fermion Yukawa couplings
$Y_f^\phi =\sqrt{2}m_fg_f^\phi/v$:
\begin{equation}
\begin{split}
g_u^h = \frac{\cos\alpha}{\sin\beta}\,,\qquad
g_u^H = \frac{\sin\alpha}{\sin\beta}\,,\qquad
g_u^A = \frac{1}{\tan\beta}\,,\\
g_d^h = -\frac{\sin\alpha}{\cos\beta}\,,\qquad
g_d^H = \frac{\cos\alpha}{\cos\beta}\,,\qquad
g_d^A = \tan\beta\,.
\label{eq:couplfermion}
\end{split}
\end{equation}
These normalized couplings are independent of the fermion
generation. 
Our calculation includes the third generation of squarks which enters
the Lagrangian in the form
\begin{align}
 \mathcal{L} \supset -(\tq_L^\dagger,\tq_R^\dagger)\mathcal{M}^2_\tq
\begin{pmatrix}\tq_L\\\tq_R\end{pmatrix}\,,
\end{align}
with the mass matrix
\begin{align}
\mathcal{M}_\tq^2=\begin{pmatrix}
M_L^2+\mass{q}^2+m_Z^2\cos(2\beta)(T_q^3-Q_qs_W^2)&
\mass{q}(A_q - \mu \kappa_q)\\
\mass{q}(A_q - \mu \kappa_q)&
M_{\tq R}^2+\mass{q}^2+m_Z^2\cos(2\beta)Q_qs_W^2
\end{pmatrix}
\label{eq:squarkmass}
\end{align}
for an arbitrary species of squarks $\tq$. This formula contains the
\susy{} soft-breaking parameters $M_L^2$, $M_{\tq R}^2$, and $A_q$. The
parameter $\mu$ determines the mass of the fermionic Higgs partners, the
Higgsinos, whereas the $Z$-boson mass $m_Z$ and the weak mixing angle
$\theta_W$ ($s_W=\sin\theta_W$) are the usual \sm{} parameters;
$\mass{q}$, $Q_q$, and $T_q^3$ are the mass, the electric, and the weak
charge of the corresponding quark $q$, respectively. It is
$\kappa_b=\tan\beta=:t_\beta$ and $\kappa_t=1/t_\beta$.

The physical particle states are obtained by the diagonalization of the mass
matrix in \eqn{eq:squarkmass} which we do in
accordance with \citere{Heinemeyer:2004xw} using
\begin{align}
\begin{pmatrix}\tq_1\\ \tq_2\end{pmatrix} = U_\tq 
\begin{pmatrix}\tq_L\\ \tq_R\end{pmatrix} \qquad\text{with}\qquad
U_\tq = \begin{pmatrix}\cos\theta_\tq&\sin\theta_\tq\\
-\sin\theta_\tq&\cos\theta_\tq\end{pmatrix}\,.
\end{align}
By choosing $0\leq \theta_q< \pi$, the masses of the squarks $\tq_1$ and
$\tq_2$ are ordered $m_{\tq 1}<m_{\tq 2}$ and given by the square roots
of the eigenvalues of $\mathcal{M}_\tq^2$ in \eqn{eq:squarkmass}:
\begin{equation}
\begin{split}
m_{\tq 12}^2= &\frac{1}{2}(M_L^2+M_{\tq
  R}^2)+\mass{q}^2+\frac{1}{2}T_q^3m_Z^2\cos(2\beta) \\ &\mp
\frac{1}{2}\sqrt{(M_L^2+M_{\tq
    R}^2+m_Z^2\cos(2\beta)(T_q^3-2Q_q\sin^2\theta_W))^2
  +4\mass{q}^2(A_q-\mu \kappa_q)^2}\,.
\end{split}
\end{equation}
The entries of $\mathcal{M}_\tq^2$ can also be expressed in terms of the
mass eigenvalues and the mixing angle
\begin{align}
\mathcal{M}_\tq^2 = \begin{pmatrix}
 m_{\tq 1}^2\cos^2\theta_\tq +  m_{\tq 2}^2\sin^2\theta_\tq&
 (m_{\tq 1}^2-m_{\tq 2}^2)\sin\theta_\tq \cos\theta_\tq\\
 (m_{\tq 1}^2-m_{\tq 2}^2)\sin\theta_\tq \cos\theta_\tq&
 m_{\tq 1}^2\sin^2\theta_\tq +  m_{\tq 2}^2\cos^2\theta_\tq
\end{pmatrix}\,,
\end{align}
which implies that the mixing angle can be obtained from
\begin{align}
\sin(2\theta_\tq) = \frac{2\mass{q}(A_q - \mu\kappa_q)}{m_{\tq
    1}^2-m_{\tq 2}^2}\,.
\label{eq:thetadefine}
\end{align}
Note that \eqn{eq:thetadefine} does not uniquely define $\theta_\tq$;
a shift $\theta_\tq\rightarrow\frac{\pi}{2}-\theta_\tq$, which corresponds to 
$\sin\theta_\tq \leftrightarrow
\cos\theta_\tq$, might be in order to allow for $m_{\tq 1}< m_{\tq 2}$.
For completeness, the couplings of the squarks
to the \mssm{} Higgs bosons can be found in \ref{sec:higgssquark}.


\subsubsection{Renormalization of the (s)top sector}
Using \eqn{eq:thetadefine}, we eliminate $A_t$ from the (s)top sector
contribution of the amplitude before renormalization and express it in
terms of the on-shell parameters for the top mass $m_t$, the stop masses
$m_{\ttop 1}$ and $m_{\ttop 2}$ and the mixing angle $\theta_{\ttop}$,
defined according to Section~3.1 in \citere{Heinemeyer:2004xw}.  In
practice, the user specifies the soft-breaking parameters $M_L\equiv
M_L(\tilde t)$, $M_{\tilde t R}$, and $A_t$, as well as the on-shell
top-quark mass $\mass{t}^\OS$. Setting $\mass{t}=\mass{t}^\OS$, \sushi{}
inserts them into the mass matrix of \eqn{eq:squarkmass} whose
eigenvalues $\mstop{1}^2,\mstop{2}^2$, as well as the corresponding stop
mixing angle $\theta_{\ttop}$ are interpreted as on-shell
parameters.\footnote{This can be seen as an indirect definition of the
  renormalization scheme for $M_L$, $M_{\tilde t R}$, and $A_t$.}  Note
that in case {\tt FeynHiggs} is used, the on-shell stop masses and the
stop mixing angle are simply taken over from its output.


\subsubsection{Renormalization of the (s)bottom sector}\label{sec:rensbot}
The renormalization of the (s)bottom sector is more subtle.  At
tree-level, the soft-breaking parameter $M_L^2$ is identical for the
sbottom and stop sector due to $SU(2)_L$ symmetry.  At higher orders,
however, in the on-shell scheme we distinguish $M_L^2(\ttop)$ and
$M_L^2(\tbot)$ similar to \citere{Yamada:1996jf,
  Bartl:1997yd,Bartl:1998xp,Eberl:1999he} by
\begin{align}
M_L^2(\tbot)  = M_L^2(\ttop) +\delta M_L^2(\ttop) - \delta M_L^2(\tbot)
\equiv  M_L^2(\ttop) + \Delta M_L^2\,,
\end{align}
with the individual counterterms given by
\begin{align}
 \delta M_L^2(\tq) = \cos^2\theta_\tq \delta m_{\tq 1}^{2,\OS}
+\sin^2\theta_\tq \delta m_{\tq 2}^{2,\OS} - (m_{\tq 1}^2-m_{\tq 2}^2)
\sin(2\theta_\tq)\delta \theta_\tq^\OS - 2\mass{q}\delta \mass{q}^\OS\,,
\end{align}
where the on-shell counterterms $\delta m_{\tq 1}^{2,\OS}$, $\delta
m_{\tq 2}^{2,\OS}$, $\delta \mass{q}^\OS$, and $\delta \theta_\tq^\OS$
are analogously defined as in the top sector, see
\citere{Heinemeyer:2004xw}.

Note that the finite shift $\Delta M_L^2$ depends on
$\msbottom{1},\msbottom{2}$, and $\theta_{\tbot}$. In order to determine its
numerical value, we first calculate ``tree-level'' values for the
sbottom masses and mixing angle by
inserting the parameters
\begin{equation}
\begin{split}
M_L^2
\equiv (M_L^2(\tilde b))^\text{tree} = M_L^2(\tilde t)\,,\qquad
M_{\tilde b R}^2\,,\qquad A_b\,,\qquad \mass{b}^\OS
\label{eq:sbottompar}
\end{split}
\end{equation}
into the mass matrix (\ref{eq:squarkmass}) (setting
$\mass{b}=\mass{b}^\OS$). All parameters of \eqn{eq:sbottompar} are
input to \sushi{}, except for the on-shell bottom-quark mass
$\mass{b}^\OS$ which is determined from the input parameter
$\mbottom(\mbottom)$ as described in \sct{sec:bottommasses}.
With these tree-level sbottom masses we define the scale
\vspace{-3mm}
\begin{equation}
\begin{split}
\muD=\frac{1}{3}(\mgluino
+ \msbottom{1}+\msbottom{2})\,.
\label{eq:mud}
\end{split}
\end{equation}

As was pointed out in
\citeres{Brignole:2002bz,Heinemeyer:2004xw,Heinemeyer:2010mm,
  Degrassi:2010eu}, replacing $A_b$ in the amplitude through
\eqn{eq:thetadefine} before renormalization -- analogous to the stop
sector -- leads to potentially large corrections $\delta A_b \propto
(\alpha_s \mu^2 \tan^2 \beta) / \mgluino$.  It was therefore suggested to
use \eqn{eq:thetadefine} in order to eliminate $\mass{b}$.
The counterterm for $\mbottom$ in this scheme, denoted $\delta \mbottom^\text{dep}$, is
obtained from \eqn{eq:thetadefine}:
\begin{align}
\delta \mbottom^\text{dep} &= 2 \mbottom \cot(2 \theta_{\tbot}) \delta\theta_{\tbot} - \frac{2
  \mbottom^{2}\cdot\delta A_b}{\sin(2 \theta_{\tbot})
  (\msbottom{1}^2-\msbottom{2}^2)} + \mbottom \frac{\delta \msbottom{1}^{2,\OS}
  -\delta
  \msbottom{2}^{2,\OS}}{\msbottom{1}^2-\msbottom{2}^2}\,.
\label{eq:dmbdep}
\end{align}
Here it is already implied that we always renormalize the sbottom masses
on-shell, while the renormalization of $A_b$ and $\theta_{\tbot}$ is still
unspecified.

In order to calculate the on-shell sbottom masses, we choose the
``on-shell'' renormalization of $A_b$, defined through a kinematical
condition on the $A\tbot_1
\tbot_2$-vertex~\cite{Brignole:2002bz,Heinemeyer:2004xw,Heinemeyer:2010mm,
  Degrassi:2010eu}.  The corresponding counterterm is
\begin{align}
&\delta A^\OS_b = (A_b+\mu \cot{\beta})
  \left[f(\msbottom{1}^2,\msbottom{2}^2)+f(\msbottom{2}^2,\msbottom{1}^2)-\frac{\delta
      \mbottom^\text{dep}}{\mbottom}\right]\,,
\label{eq:dabmb} 
\end{align}
with
\vspace{-3mm}
\begin{equation}
\begin{split}
&\quad f(m_{1}^2,m_{2}^2) = -\frac{\alpha_s(\muD)}{\pi}\frac{2}{3}
\left\{\left[-\frac{\mgluino}{A_b+\mu \cot{\beta}}
B_0^{\text{fin}}(m_{1}^2,\mbottom,\mgluino,\muD)\right]\right.  \\ 
& \left.\qquad+\frac{m_{1}^2}{m_{1}^2-m_{2}^2}
\left[4+2 \log\frac{\muR^2}{m_{1}^2}-(1-\frac{\mgluino^2}{m_{1}^2}
-\frac{\mbottom^{2}}{m_{1}^2})\cdot B_0^{\text{fin}}(m_{1}^2,\mbottom,\mgluino,\muD)\right]\right\}\,,
\end{split}
\end{equation}
where the function $B_0^{\text{fin}}$ can be taken from Eq.\,(B.8) in
\citere{Harlander:2004tp} replacing $\muR\rightarrow \muD$.  
Solving \eqn{eq:dmbdep} (with $\delta A_b=\delta A_b^\OS$) and
\eqn{eq:dabmb} for $\delta A^\OS_b$, we obtain
\begin{align}\nonumber
\delta A^\OS_b &= (A_b \sin\beta+\mu \cos\beta)
\left\{-(\msbottom{1}^2-\msbottom{2}^2) \sin (2 \theta_{\tbot})
\left[f(\msbottom{1}^2,\msbottom{2}^2)+f(\msbottom{2}^2,\msbottom{1}^2)\right]\right.
\\ &\qquad\qquad +\left.(\delta \msbottom{1}^{2,\OS}-\delta
\msbottom{2}^{2,\OS}) \sin (2 \theta_{\tbot})+2 \delta\theta_{\tbot} (\msbottom{1}^2
- \msbottom{2}^2) \cos (2 \theta_{\tbot})\right\} \label{eq:dabtb}
\\ \nonumber &\quad\cdot\left[2 (\sin \beta (A_b
  \mbottom-(\msbottom{1}^2-\msbottom{2}^2) \sin \theta_{\tbot} \cos \theta_{\tbot})+\mbottom
  \mu \cos \beta)\right]^{-1}\,,
\end{align}
which in turn allows to calculate $\delta \mbottom^\text{dep}$. Note that 
one is still free to choose the renormalization condition for
$\theta_{\tbot}$.

The on-shell sbottom masses are finally obtained as follows: We
calculate the counterterm $\delta \mbottom^\text{dep}$ from \eqn{eq:dmbdep}
using $\delta A_b=\delta A_b^\OS$ and
$\delta\theta_{\tbot}=\delta\theta_{\tbot}^\OS$, the ``tree-level'' sbottom masses
and mixing angle and the on-shell bottom mass. This yields a
numerical value for $\Delta M_L^2$ and similarly
for $\mbottom^\text{dep}=\mbottom^{\drbar}(\muD)-\delta
\mbottom^\text{dep}+\delta \mbottom^{\drbar}$,
where the $\drbar$ mass is calculated in the~\sm{}.
Diagonalizing the mass matrix $\mathcal{M}_\tbot^2$ in
\eqn{eq:squarkmass} with $\mbottom=\mbottom^\text{dep}$, we thus obtain on-shell
sbottom masses $m_{\tbot 1}$ and $m_{\tbot 2}$ and the on-shell sbottom
mixing angle~$\theta_{\tbot}$.
In case {\tt FeynHiggs} is called for the calculation of the \mssm{}
Higgs masses, we take $\mbottom^\text{dep}$ as well as $\Delta M_L^2$ from
this program which, when inserted into the sbottom mass matrix
\eqn{eq:squarkmass}, results in on-shell sbottom masses and mixing
angle consistent with the values given by {\tt FeynHiggs}
itself.

After the on-shell sbottom masses have been determined, \sushi{} allows
for a change between various schemes regarding the renormalization of
$\mbottom$, $A_b$, and $\theta_\tbot$ in the sbottom contribution to the
gluon-fusion amplitude.  Note, however, that either $\mbottom$ or $A_b$ is required
to be a dependent quantity; a dependent renormalization of
the sbottom mixing angle is not offered as option.  The possible choices
are summarized in \tab{tab:sbottomrenorm}.
Numerical differences in the
cross sections between the various renormalization schemes and their
implications will be investigated in a separate publication.

\begin{table}[htb]
\begin{center}
\begin{tabular}{| c || c | c | c |}
\hline
\multicolumn{4}{|c|}{Scheme choices}\\\hline\hline
$\mbottom$      	& \red{dep.}	& OS & $\drbar$ \\
$A_b$      	& dep.	& \red{OS} & $\drbar$ \\
$\theta_\tbot$	& 	& \red{OS} & $\drbar$ \\\hline
\end{tabular}
\end{center}
\caption{Available renormalization schemes for the sbottom sector.
The default option is marked in red.}
\label{tab:sbottomrenorm}
\end{table}

To summarize, apart from $\mu$, $\tan\beta$ and the \sm{} parameters,
the input parameters of \sushi{} that determine the
squark sectors are:
\begin{itemize}
\item $M_L^2$, $M_{\tilde t R}^2$, $A_t$, $\mass{t}^\OS$ which directly
  determine the on-shell stop masses and mixing angle through
  diagonalization of \eqn{eq:squarkmass}
\item $M_{\tilde b R}^2$, $A_b$, $\mbottom(\mbottom)$, where
  $A_b$ is understood as renormalized according to \eqn{eq:dabmb}.
\end{itemize}
Switching between renormalization schemes changes both the counterterms
to the amplitude as well as the numerical values of the parameters
$\mbottom$, $\msbottom{1}$, $\msbottom{2}$, $A_b$, and $\theta_{\tbot}$,
of course.  We remark that changing the renormalization scheme in the
sbottom sector affects $\mbottom$ only in the Higgs-sbottom couplings;
the renormalization of the bottom mass occurring in the Higgs-bottom
Yukawa coupling is independent of that (see \sct{sec:bottommasses}). For
the bottom mass occurring in internal propagators (rather than in
couplings), \sushi{} always uses the on-shell value.

The switching between the different renormalization schemes is done at
the renormalization scale $\muR$ to guarantee the same coupling strength
$\alpha_s(\muR)$ for the \nlo{} counterterms and the cross section
itself. In case \sushi{} makes use of the formulas in
\citeres{Degrassi:2010eu,Degrassi:2011vq}, the counterterms at \nlo{}
are expanded to the correct order in the bottom mass to match the
expanded \nlo{} amplitudes.



\subsection{Bottom mass calculation/Resummation of $\tan\beta$-enhanced contributions}
\label{sec:bottommasses}

The bottom-quark mass in the input files of \sushi{} is inserted in the
$\msbar$ scheme $\mbottom(\mbottom)$.  Together with the input value of
$\alpha_s(m_Z)$, we calculate $\alpha_s(\mbottom(\mbottom))$ by $4$-loop
running with $5$ active flavors.  Using Eq.~(13) of
\citere{Melnikov:2000qh} at $3$-loop level (see also
\citere{Chetyrkin:1999qi}), $\mbottom(\mbottom)$ is transformed into its
on-shell value $\mbottom^{\OS}$. 

In the \sm{}, the bottom Yukawa coupling $Y_b=\sqrt{2}\mbottom^Y/v$ can be
chosen $\mbottom^Y=\mbottom^{\OS}$ or alternatively $\mbottom^Y=\mbottom^{\msbar}(\mu_b)$,
where $\mu_b\in \lbrace \mbottom,\muR\rbrace$; $\muR$ denotes the
renormalization scale.  As indicated above, the bottom mass entering the
internal propagators is always set to the on-shell mass $\mbottom^{\OS}$ in
\sushi{}.  In the \mssm{}, \sushi{} offers various options regarding the choice
of the Higgs-bottom Yukawa coupling $Y_b^\phi=\sqrt{2} \mbottom^{Y,\phi}g_f^\phi/v$:
{\allowdisplaybreaks
\begin{subequations}
\begin{align}
 & &\text{ on-shell coupling: } &\mbottom^{Y,\phi} =  \mbottom^{\OS} \label{eq:oscoupling}\\
 & &\text{ basic resummation: } &\mbottom^{Y,\phi} = \frac{\mbottom^{\OS}}{1+\Delta_b} \label{eq:naiveres}\\
 & &\text{ full resummation $h$: } &\mbottom^{Y,h} = \frac{\mbottom^{\OS}}{1+\Delta_b}\left(1-\Delta_b\frac{\cot\alpha}{\tan\beta}\right)\\
 & &\text{ $H$: } &\mbottom^{Y,H} = \frac{\mbottom^{\OS}}{1+\Delta_b}\left(1+\Delta_b\frac{\tan\alpha}{\tan\beta}\right)\\
 & &\text{ $A$: } &\mbottom^{Y,A} = \frac{\mbottom^{\OS}}{1+\Delta_b}\left(1-\Delta_b\frac{1}{\tan^2\beta}\right)\\
 & &\text{ Running coupling: } &\mbottom^{Y,\phi} =  \frac{\mbottom^{\msbar}(\mu_b)}{1+\Delta_b} \label{eq:msbarcoupling}
\end{align}
\end{subequations}

\begin{figure}[ht]
\begin{minipage}{0.47\textwidth}
\begin{center}
\includegraphics[width=0.7\textwidth]{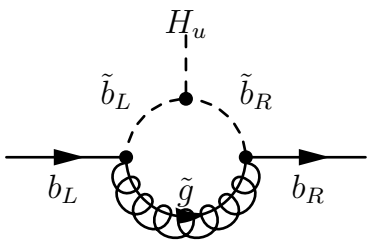}
\end{center}
\vspace{-1cm}
\caption{Feynman diagram inducing an effective coupling of the bottom-quarks to $H_u$. The coupling is
proportional to $\mu t_\beta$ and can be resummed in $\Delta_b$.
}
\label{fig:tanbresum}
\end{minipage}\hfill
\begin{minipage}{0.53\textwidth}
Herein, $\Delta_b$ resums higher order sbottom contributions as shown in
\fig{fig:tanbresum}~\cite{Banks:1987iu,Hall:1993gn,Hempfling:1993kv,
Carena:1994bv,Carena:1999py,Carena:2000uj}, for example,
calculated at the scale $\mu_\text{d}$, defined in \eqn{eq:mud}.
The exact formula is given by:
\begin{align}\label{eq:deltab}
\Delta_b = \frac{2}{3\pi}\alpha_s(\mu_\text{d})\mgluino\mu t_\beta I(\msbottom{1}^2,\msbottom{2}^2,\mgluino^2)\\
I(a,b,c)=\frac{ab\ln(\tfrac{a}{b})+bc\ln(\tfrac{b}{c})+ca\ln(\tfrac{c}{a})}{(a-b)(b-c)(a-c)}\label{eq:ifunc}
\end{align}
\end{minipage}
\end{figure}
If the input values for \sushi{} are determined by {\tt FeynHiggs},
\sushi{} allows to use the value of $\Delta_b$ as given in the output of
this program which contains also the electro-weak contributions from
neutralinos and charginos in accordance with \citere{Hofer:2009xb};
two-loop corrections to
$\Delta_b$~\cite{Noth:2010jy,Noth:2008tw,Bauer:2008bj} are not yet
included.  For the running coupling in \eqn{eq:msbarcoupling}, the scale
$\mu_b$ can be set to $\mbottom$ or $\muR$. The numerical differences
between the various schemes will be discussed in a forthcoming
publication.



\section{Cross section for gluon fusion}

As indicated in the Introduction, the most important production channel
in the \sm{} and for moderate values of $t_\beta$ in the \mssm{} is
gluon fusion. 
After quoting the well-known results for the \lo{} cross
section, our implementation of the \nlo{} contributions
is explained. A discussion of \nnlo{} and electro-weak contributions follows.

Using the notation of \citere{Spira:1995rr}, the hadronic
cross section for $\phi\in\lbrace h,H,A \rbrace$ at \nlo{} \qcd{} can be
written as follows:
\begin{align}
\sigma(pp\rightarrow \phi+X)
=\sigma_0^\phi\left[1+C^\phi\frac{\alpha_s}{\pi}\right]\tau_\phi
\frac{d\mathcal{L}^{gg}}{d\tau_\phi}
+\Delta \sigma_{gg}^\phi + \Delta \sigma_{gq}^\phi + \Delta \sigma_{q\overline{q}}^\phi\,,
\label{eq:crosssection}
\end{align}
where $\tau_\phi = \mass{\phi}^2/s$, with the hadronic center-of-mass
energy $s$. The factor $\sigma_0^\phi$ is determined by the \lo{} cross
section, $C^\phi$ arises from \nlo{} terms in the partonic cross section
that are singular as $\hat s\to \mass{\phi}^2$ ($\hat s$ is the partonic
center-of-mass energy), and
\begin{align}
 \frac{d\mathcal{L}^{gg}}{d\tau} = \int_\tau^1 \frac{dx}{x} g(x)g(\tau/x)
\end{align}
is the gluon-gluon luminosity. The quantities $\Delta\sigma_{gg}^\phi$,
$\Delta \sigma_{gq}^\phi$, and $\Delta \sigma_{q\overline{q}}^\phi$
comprise the terms that are regular as $\hat s\to \mass{\phi}^2$ in the
partonic cross section, arising from $gg$, $gq$ and $q\overline{q}$
scattering, respectively. Loosely speaking, $C^\phi$ is due to the
virtual, while the $\Delta\sigma_{ij}^\phi$ are due to the real radiation
contributions. The latter are implemented in \sushi{} by expressing them
in terms of Passarino-Veltman functions~\cite{Passarino:1978jh}, see
\citere{Harlander:2010wr}.  The strong coupling $\alpha_s$ for the cross
section calculations is renormalized in standard \qcd{} with five active
quark flavors.


\subsection{\lo{} cross section}\label{sec:lo}
Taking into account the third generation of quarks (and squarks in the
\mssm{}), the normalization factor in \eqn{eq:crosssection} can be
written in the form
\begin{align}
 \sigma_0^\phi=\frac{G_F\alpha_s^2(\muR)}{288\sqrt{2}\pi}|\mathcal{A}^{\phi,(0)}|^2\,,
\label{eq:sig0}
\end{align}
with Fermi's constant $G_F$. The amplitude $\mathcal{A}$ for is given by
\begin{align}
&\mathcal{A}^{\phi,(0)} = \sum_{q\in\lbrace t,b\rbrace}
  \left(a_q^{\phi,(0)}+\tilde{a}_q^{\phi,(0)}\right)\,,
  \label{eq:amp0}
\end{align}
with the individual contributions for $\phi\in\{h,H\}$
\begin{align}
a_q^{\phi,(0)}&=g_q^\phi\frac{3\tau_q^\phi}{2}(1+(1-\tau_q^\phi)f(\tau_q^\phi)),\qquad
\tilde{a}_q^{\phi,(0)}=-\frac{3\tau_q^\phi}{8}\sum_{i=1}^2
g_{\tq,ii}^\phi(1-\tau_{\tq i}^\phi f(\tau_{\tq i}^\phi))\,,
\end{align}
using the notation
\begin{align}
\tau_q^\phi = \frac{4\mass{q}^2}{\mass{\phi}^2}\,,\qquad
\tau_{\tq i}^\phi = \frac{4m_{\tq i}^2}{\mass{\phi}^2}
\end{align}
and the function
\begin{align}
 f(\tau) = \left\lbrace \begin{matrix}
 \arcsin^2\frac{1}{\sqrt{\tau}}&\tau\geq 1\\
-\frac{1}{4}\left(\log\frac{1+\sqrt{1-\tau}}{1-\sqrt{1-\tau}}-i\pi\right)^2&\tau <1
\end{matrix}\right.\,.
\end{align}
For the {\abbrev CP}-odd Higgs $\phi=A$, the squarks do not contribute
at \lo{}, i.e. $\tilde{a}_q^{A,(0)}=0$, and the quark contribution can be
written in the form
\begin{align}
 a_q^{A,(0)}=g_q^A\frac{3\tau_q^A}{2}\tau_q^Af(\tau_q^A)\,.
\end{align}
The couplings $g_f^\phi$ of the Higgs $\phi$ to the quarks can be taken
from \eqn{eq:couplfermion}, the couplings $g_{\tq,ij}^\phi$ to the
squarks from \ref{sec:higgssquark}. Needless to say, in the \sm{}, the
squark couplings have to be set to zero and the quark couplings to
$g_q^\phi=1$.


\subsection{\nlo{} virtual contributions}\label{sec:nlo}
As indicated above, the coefficient $C^\phi$ contains the virtual
corrections to the $gg$ process and is regularized by the infrared
singular part; moreover, it includes the counterterms to \lo{}
quantities. We write it as
\begin{align}
 C^\phi=2\Re\left[\frac{\mathcal{A}^{\phi,(1)}}{\mathcal{A}^{\phi,(0)}_\infty}\right]
+\pi^2+\beta_0\log\left(\frac{\muR^2}{\muF^2}\right)\,,
\end{align}
where $\beta_0=11/2-n_f/3$ with $n_f=5$; $\muF$ and $\muR$ denote the
factorization and the renormalization scale, respectively. The \nlo{}
amplitude $\mathcal{A}^{\phi,(1)}$ and the \lo{} amplitude
$\mathcal{A}^{\phi,(0)}_\infty$ in the limit of large stop and sbottom
masses are given by
\begin{align}\label{eq:NLOamp}
\mathcal{A}^{\phi,(1)}=\sum_{q\in\lbrace t,b\rbrace}(a_q^{\phi,(1)}+\tilde{a}_q^{\phi,(1)}),\qquad
\mathcal{A}_\infty^{\phi,(0)} = \sum_{q\in\lbrace t,b\rbrace}
\left(a_q^{\phi,(0)}+\frac{\tau_q^\phi}{8}\sum_{i=1}^2\frac{g_{\tq,ii}^\phi}{\tau_{\tq i}^\phi}\right)\,.
\end{align}
Available results for the \nlo{} contributions have been discussed in
the Introduction.
In \sushi{}, we use the analytic formulas of \citere{Harlander:2005rq}
for the quark-induced terms $a_q^{\phi,(1)}$. The purely squark-induced
terms (see \fig{fig:nlocont}\,(a), for example) need to be considered in
combination with the mixed quark/squark/gluino diagrams (two examples
are shown in \fig{fig:nlocont}) in order to preserve supersymmetry,
resulting in the coefficient $\tilde a_q^{\phi,(1)}$. \sushi{}
implements expansions for these amplitudes in two limits:
\begin{itemize}
\item $\mass{\phi} \ll
  \mass{q},\msquark{1},\msquark{2},\mgluino$~\cite{Harlander:2004tp,%
    Harlander:2003bb,Degrassi:2008zj} which is valid through
  $\mass{\phi}<\min(2\mass{q},2\msquark{},\msquark{}+\mass{q}+\mgluino)$
  and thus applies to the top-stop sector as long as $\phi$ is not too
  heavy. \sushi{} incorporates the publicly available program {\tt
    evalcsusy.f}~\cite{evalcsusy} in order to use this result for the light
  Higgs $h$.
\item $\mass{\phi},\mass{q} \ll
  \msquark{1},\msquark{2},\mgluino$~\cite{Harlander:2010wr,Degrassi:2010eu,%
    Degrassi:2012vt,Degrassi:2011vq} which holds through $\mass{\phi} <
  \min(2\msquark{},\msquark{}+\mgluino)$, and thus applies to the
  bottom-sbottom sector, as well as to the top-stop sector when $\phi$
  is heavy; \sushi{} uses the formulas of
  \citeres{Degrassi:2012vt,Degrassi:2011vq} in these cases.
\end{itemize}
Note that both expansions hold only as long as the Higgs mass is not
much heavier than the typical \susy{} mass. For larger Higgs masses,
only the fully numerical result of \citere{Anastasiou:2008rm} is known
so far.

\begin{figure}[ht]
\begin{center}
\begin{tabular}{ccc}
\includegraphics[width=0.3\textwidth]{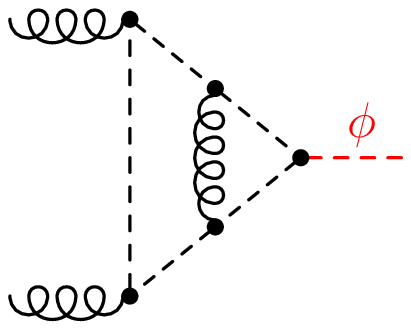}  &
\includegraphics[width=0.3\textwidth]{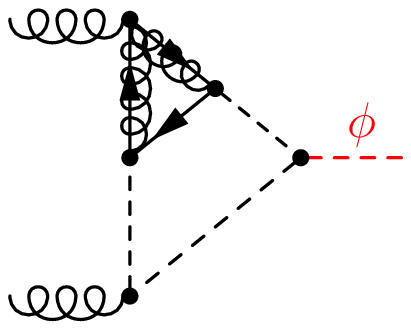} &
\includegraphics[width=0.3\textwidth]{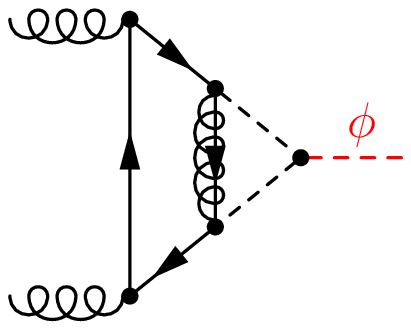} \\
(a) & (b) & (c)
\end{tabular}
\end{center}
\vspace{-0.8cm}
\caption{Example Feynman diagrams contributing to
  $\tilde{a}_q^{\phi,(1)}$: gluon-squark contribution (left),
  gluino-quark-squark contribution (middle) and gluino-quark-squark
  contribution (right). The latter is partially resummed by the usage of
  $\Delta_b$ from \eqn{eq:deltab}.  }
\label{fig:nlocont}
\end{figure}


\subsection{\nnlo{} corrections}
\label{sec:nnlocorrections}

While the implementation of \nlo{} corrections allows for the evaluation
of inclusive and exclusive quantities, \nnlo{} corrections are available
in \sushi{} only for the total inclusive cross section.  In addition,
only the top-(s)quark induced gluon-Higgs coupling is taken into account
at \nnlo{}; the top-squark induced one only approximately:
\begin{align}
\sigma_{gg\phi,\text{\nnlo}}^\text{\mssm{}} =
\sigma^\text{\mssm{}}_{gg\phi,\text{\nlo{}}} +
(\sigma^{t}_{gg\phi,\text{\nnlo{}}}
-\sigma^{t}_{gg\phi,\text{\nlo{}}})\,,
\label{eq:ggh@nnlo}
\end{align}
where, on the right-hand side, the \nnlo{} term is evaluated with the
\nnlo{} \pdf{}s, while the \nlo{} terms are evaluated as usual with
\nlo{} \pdf{}s.

The \nbnlo{} top-(s)quark contributions to the cross sections
$\sigma^{t}_{gg\phi,\nbnlo{}}$ are calculated with the help of the
programs {\tt ggh@nnlo} \cite{ggh@nnlo} and
{\tt evalcsusy} \cite{evalcsusy} which work in the effective theory
approach of heavy top (s)quarks. The \nnlo{} top-squark (and mixed
top/stop/gluino) effects, available for $\phi=h$ only, have been
evaluated~\cite{Pak:2010cu,Pak:2012xr} by applying the limit
$\mass{\phi} \ll \mass{q},\msquark{1},\msquark{2},\mgluino$. The result
has been implemented in a computer code that involves {\tt Mathematica}
and a number of other programs. \sushi{} includes an approximation of
these \nnlo{} effects according to
Ref.\,\cite{Harlander:2003kf}\footnote{The \nnlo{} term of the
  normalized Wilson coefficient ($\kappa_2$ in the notation of
  \citere{Harlander:2003kf}) is replaced by its \sm{} value in
  \sushi{}.}.


\subsection{Electro-weak corrections}
\label{sec:ewcontributions}

The full \nlo{} electro-weak ({\abbrev EW}) corrections are known only in the \sm{}
\cite{Actis:2008ug}. It has been suggested to assume complete
factorization of \qcd{} and {\abbrev EW}
effects~\cite{Anastasiou:2008tj}, thus writing
\begin{equation}
\begin{split}
\sigma_{ggH,\text{\nnlo,{\abbrev EW}}}^{\text{\sm},t} = (1+\delta_\text{\abbrev EW})
\sigma_{ggH,\text{\nnlo}}^{\text{\sm},t}\,.
\label{eq:fullfac}
\end{split}
\end{equation}

For the {\abbrev CP}-even Higgs bosons in the \mssm{},
a formula based on \eqn{eq:ggh@nnlo} and
\eqn{eq:fullfac} has been used for the combination of \qcd{} and
electro-weak corrections in \citere{Harlander:2010wr}:
\begin{align}
\sigma_{gg\phi,\text{\nnlo},\text{\abbrev EW}}^\text{\mssm{}} =
\sigma_{gg\phi,\text{\nlo{}}}^\text{\mssm{}} +
(1+\delta_{\text{\abbrev
      EW}})\sigma^{t}_{gg\phi,\text{\nnlo{}}}
  -\sigma^{t}_{gg\phi,\text{\nlo{}}}\,,
\label{eq:ggh@nnloew}
\end{align}
or, at \nlo{} precision,
\begin{align}
\sigma_{gg\phi,\text{\nlo},\text{\abbrev EW}}^\text{\mssm{}} =
\sigma_{gg\phi,\text{\nlo{}}}^\text{\mssm{}} +
\delta_{\text{\abbrev
      EW}}\sigma^{t}_{gg\phi,\text{\nlo{}}}\,.
\end{align}

Alternatively, it has been suggested in \citere{Bagnaschi:2011tu} to use
the \sm{} electro-weak corrections due to light
quarks only~\cite{Aglietti:2004nj,Bonciani:2010ms}.  Following
\citere{Bagnaschi:2011tu}, we define the correction factor
\begin{align}
 \delta_{\text{EW}}^\text{lf} = \frac{\alpha_{\text{EM}}}{\pi}
 \frac{2\text{Re}\left(\mathcal{A}^{\phi,(0)}\mathcal{A}^{\phi,\text{EW}}\right)}{|\mathcal{A}^{\phi,(0)}|^2}\,,
\end{align}
where $\mathcal{A}^{\phi,(0)}$ denotes the complete \lo{} amplitude
including quark and squark diagrams, see \eqn{eq:amp0}, and the
electro-weak amplitude is given by~\cite{Bonciani:2010ms}
\begin{align}
&\mathcal{A}^{\phi,\text{EW}} =
- \frac{3}{8}
\frac{x_W}{s_W^2}
\left[\frac{2}{c_W^2}\left(\frac{5}{4}-\frac{7}{3}s_W^2+
\frac{22}{9}s_W^4\right)A_1[x_Z] + 4A_1[x_W]\right]g_V^\phi\,,
\end{align}
with 
\begin{equation}
\begin{split}
x_V = \frac{1}{\mphi^2}\left(m_V-i\frac{\Gamma_V}{2}\right)^2\,,
\qquad V\in\{W,Z\}\,,
\end{split}
\end{equation}
the electro-magnetic coupling $\alpha_{\text{EM}}$, and
$s_W=\sin\theta_W=(1-c_W^2)^{1/2}$ the sine of the weak mixing angle.
Supersymmetry enters through the relative couplings $g_V^\phi$ given by
\begin{align}
g_V^h = \sin(\beta-\alpha),\qquad g_V^A=0,\qquad g_V^H =
\cos(\beta-\alpha)\,. 
\end{align}

The function $A_1[x]$ can be found in
\citeres{Aglietti:2004nj,Bonciani:2010ms}.  Since its numerical
evaluation is rather involved, \sushi{} implements $\delta_\text{\abbrev
  EW}^\text{lf}$ (and $\delta_\text{\abbrev EW}$) in terms of an
interpolation grid in $m_\phi$ (and $m_t$), using fixed values for the
gauge boson masses and widths, as well as for the weak mixing angle
($m_W=80.385$\,GeV, $\Gamma_W=2.085$\,GeV, $\sin^2\theta_W=0.22295$,
$m_Z=91.1876$\,GeV, $\Gamma_Z=2.4952$\,GeV~\cite{Beringer:1900zz}); the
input values to \sushi{} for these parameters are ignored in the
evaluation of the electro-weak corrections.  The electro-weak correction
factor due to light quarks $\delta_{\text{EW}}^\text{lf}$ multiplies the
\nlo{} \mssm{} cross section, while the \nnlo{} \qcd{} effects are
simply added as in \eqn{eq:ggh@nnlo}:
\begin{align}
\sigma_{gg\phi}^\text{\mssm{}} =
\sigma_{gg\phi,\text{\nlo{}}}^\text{\mssm{}}(1+\delta_{\text{\abbrev
    EW}}^\text{lf}) +
\sigma^{t}_{gg\phi,\text{\nnlo{}}}-\sigma^{t}_{gg\phi,\text{\nlo{}}}\,.
\label{eq:ewlf}
\end{align}
\sushi{} leaves it up to the user to decide whether to use \eqn{eq:ggh@nnloew}
or \eqn{eq:ewlf} in order to include the electro-weak
corrections. For a \sm{}-like Higgs and $m_\phi<2m_t$, both approaches
lead to comparable \nlo{} results. The {\abbrev EW} corrections for a
{\abbrev CP}-odd Higgs are not known and thus not included.



\section{Cross section for bottom-quark annihilation}
In supersymmetric theories, where the Higgs coupling to bottom-quarks
can be enhanced by $\tan\beta$, associated production $(b\overline{b})
\phi + X$ can be similarly or even more important than gluon fusion. Two
theoretical approaches have been pursued for the theoretical description
of this process: In the four-flavor scheme (\fs{4}), the relevant
production processes at lowest order \qcd{} are $gg\rightarrow
(b\overline{b})\phi$ (see \fig{fig:bbh45FS}\,(a)) and quark-antiquark
annihilation $q\overline{q}\rightarrow (b\overline{b})\phi$
\cite{Raitio:1978pt,Ng:1983jm,Kunszt:1984ri}.  However, when integrating
over all final-state bottom-quark momenta, potentially large logarithms
$\ln \mbottom/\mphi$ occur.  They can be resummed by the introduction of
bottom-quark \pdf{}s, which defines the five-flavor scheme (\fs{5})
\cite{Barnett:1987jw,Dicus:1988cx}. The \lo{} process in this latter
scheme is bottom-quark annihilation $b\overline{b}\rightarrow \phi$ for
which the lowest-order Feynman diagram is shown in
\fig{fig:bbh45FS}\,(b).

\begin{figure}[ht]
\begin{center}
\begin{tabular}{cc}
\includegraphics[width=0.28\textwidth]{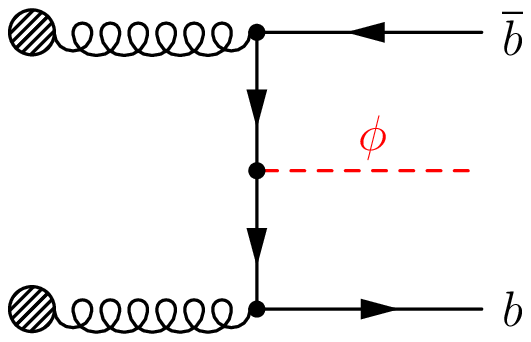}\hspace{1cm}
&\includegraphics[width=0.26\textwidth]{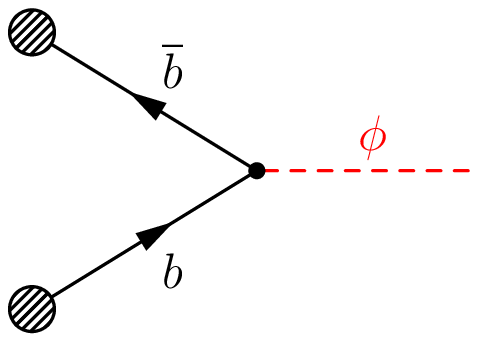}\\
(a) & (b)
\end{tabular}
\end{center}
\vspace{-0.6cm}
\caption{Feynman diagrams showing the associated production
process $gg\rightarrow (b\overline{b})\phi$ in the $4$FS (left)
and bottom-quark annihilation in the $5$FS (right).
}
\label{fig:bbh45FS}
\end{figure}

\sushi{} implements results for associated $b\bar b\phi$-production in
the \fs{5}.  For the inclusive cross section, it links the program {\tt
  bbh@nnlo}~\cite{Harlander:2003ai} in order to obtain the \nnlo{}
\qcd{} prediction $\sigma_{bbH}^\text{\sm{}}$ using $\mbottom^{\msbar}(\muR)$
for the bottom Yukawa coupling.  This is then re-weighted by the
corresponding resummed \susy{} coupling
$\tilde{g}_b^h$~\cite{Carena:1999py,Guasch:2003cv} as follows
\begin{align}\label{eq:bbhreweighting}
&\sigma_{bb\phi}^{\text{\mssm{}}} = \sigma_{bbH}^{\text{\sm{}}}\cdot
(\tilde{g}_b^\phi)^2 \qquad\text{with}\qquad \tilde{g}_b^h =
\frac{g_b^h}{1+\Delta_b}\left(1-
\Delta_b\frac{\cot\alpha}{\tan\beta}\right)\,,\\
&\quad\tilde{g}_b^H=
\frac{g_b^H}{1+\Delta_b}\left(1+\Delta_b\frac{\tan\alpha}{\tan\beta}\right)\,,
\quad \tilde{g}_b^A=
\frac{g_b^A}{1+\Delta_b}\left(1-\Delta_b\frac{1}{\tan^2\beta}\right)\,,
\end{align}
where the $g_b^\phi$ are given in \eqn{eq:couplfermion} and
$\Delta_b$ is determined as described in \sct{sec:bottommasses}.

For differential cross sections due to bottom-quark annihilation,
\sushi{} includes the \nlo{} virtual corrections for $b\bar b\to \phi$
and combines them with the \lo{} real-radiation processes $b\bar b\to
g\phi$ and $bg\to b\phi$ using dipole subtraction.\footnote{We are
grateful to M.\,Wiesemann for providing us with the corresponding {\tt
Fortran} routines which entered the studies presented in
\citeres{Harlander:2011fx,Harlander:2010cz}.}  Similar to the fully
inclusive case, we multiply with the resummed \susy{} couplings to
obtain \mssm{} cross sections.

\section{Differential cross sections}
Apart from the total inclusive cross sections due to gluon fusion and
bottom-quark annihilation, \sushi{} also allows for the computation of
differential cross sections in these processes. In particular, one may
apply upper and lower cuts on the Higgs transverse momentum $p_T$, its
rapidity $y$ or its pseudo-rapidity $\eta$, where
\begin{align}
\eta=-\ln\left(\tan\frac{\theta}{2}\right)=\frac{1}{2}
\left(\frac{|\vec{p}|+p_L}{|\vec{p}|-p_L}\right),
\qquad
y=\frac{1}{2}\left(\frac{E+p_L}{E-p_L}\right)\,.
\end{align}
Here, $\vec{p}=\vec{p}_T+\vec{p}_L$ is the 3-momentum of the Higgs
boson, $p_L$ the longitudinal component, $E$ the Higgs boson's energy,
and $\theta$ the scattering angle (all in the hadronic reference frame).
For gluon fusion, \sushi{} also provides the differential quantities
$d\sigma/dp_T$, $d\sigma/dy$, and $d^2\sigma/(dp_Tdy)$ (or,
alternatively, $d\eta$ instead of $dy$).  We add that, since the
distribution in $y$ and $\eta$ is symmetric, minimal and maximal values
for $y$ are understood as $0 \leq y_{\text{min}}\leq |y| \leq
y_{\text{max}}$ (and similarly for $\eta$).  In order to get reliable
results, the precision for the numerical integration in \sushi{} should
be set to a higher value for differential quantities than for inclusive
cross sections.

Note that at \lo{}, i.e., $\order{\alpha_s^2}$ for gluon fusion and
$\order{\alpha_s^0}$ for bottom-quark annihilation, the Higgs
transverse momentum is always $p_T=0$.  \sushi{} provides results for
non-inclusive quantities through \nlo{}, i.e., $\order{\alpha_s^3}$ for
gluon fusion and $\order{\alpha_s}$ for bottom-quark annihilation.  Let
us also add that $p_T$-cuts or $p_T$-distributions should not be too low
($p_T/m_\phi\gtrsim 0.1$), since otherwise potentially large logarithms
may spoil the perturbative convergence of the fixed-order results
implemented in \sushi{}. For the resummation of such terms in Higgs
production, see Refs.\,\cite{Bozzi:2005wk,Mantler:2012bj}, for example.


\section{The program \sushi{}}
\label{sec:program}

This section describes the most important technical details of the
program \sushi{}, including its installation and usage.

\subsection{Workflow}

The workflow of \sushi{} is depicted in \fig{fig:workflow}.  The input
is controlled by a single input file whose format is {\abbrev
  SLHA}-inspired \cite{Skands:2003cj,Allanach:2008qq}.  In case of the
\mssm{}, the user specifies whether the Higgs mass is calculated by {\tt
  FeynHiggs} or provided by the user himself.  After the initialization
of internal parameters which are derived from the input data, \sushi{}
transforms them to the specified renormalization scheme and determines
the resummation of $\tan\beta$-enhanced terms in the bottom Yukawa
coupling, see \scts{sec:rensbot} and \ref{sec:bottommasses}.
Afterwards, the gluon fusion and bottom-quark annihilation cross
sections are calculated up to the desired perturbative order.  The
\nnlo{} top-(s)quark induced and the electro-weak contributions for
gluon fusion are taken into account only for the inclusive cross
section. Not shown in the workflow is the link to {\tt LHAPDF} which
occurs at various stages of the internal calculation.  The output is
printed to the screen and written to an output file which follows the
same format as the input file. Details concerning the in- and output
files are given in \sct{sec:inout}.

\begin{figure}[ht]
\begin{tikzpicture}[]
\path (0.0,0.0) node[below,boxred]{%
\begin{minipage}{0.4\textwidth}
\centering  \sushi{} input file in SLHA-style
\end{minipage}};

\path (0.0,-2.0) node[below,boxblue]{%
\begin{minipage}{0.30\textwidth}
\centering \sushi{} initialization
\end{minipage}};

\path[->,ultra thick,blue] (3.5,-0.5) edge [bend left=70]
node[midway,right,black,boxgreen]{
\begin{minipage}{0.22\textwidth}
\centering \susy{}: Higgs mass by {\tt FeynHiggs}
\end{minipage}
} (2.65,-2.3);

\path[->,ultra thick,blue] (-3.5,-0.5) edge [bend right=70]
node[midway,left,black,boxblue]{
\begin{minipage}{0.22\textwidth}
\centering \sm{}, \susy{}: Higgs mass as input
\end{minipage}
} (-2.65,-2.3);

\path[->,ultra thick,blue] (0.0,-2.7) edge [] (0.0,-3.2);

\path (0.0,-3.2) node[below,boxblue]{%
\begin{minipage}{0.7\textwidth}
\centering Transformation to specified renormalization scheme;
calculation of couplings (including resummation)
\end{minipage}};


\path[->,ultra thick,blue] (-2.5,-4.5) edge [bend right=40]
node[midway,left,black,boxblue]{
\begin{minipage}{0.22\textwidth}
\centering Total cross section
\end{minipage}
} (-4.5,-7);

\path (-4.5,-6.5) node[below,boxblue]{%
\begin{minipage}{0.39\textwidth}
\centering Calculation of ggh/bbh at \nbnblo{}
\end{minipage}};

\path (0,-5.3) node[below,boxgreen]{%
\begin{minipage}{0.3\textwidth}
\centering {\tt bbh@nnlo}, bbh diff.
\end{minipage}};

\path[->,ultra thick,blue] (-2.7,-5.6) edge [bend right=40] (-3.5,-6.5);
\path[->,ultra thick,blue] (2.7,-5.6) edge [bend left=40] (3.5,-6.5);

\path (0,-7.5) node[below,boxgreen]{%
\begin{minipage}{0.4\textwidth}
\centering {\tt ggh@nnlo}, electro-weak contr.
\end{minipage}};

\path[->,ultra thick,blue] (-4.5,-7.25) edge [] (-4.5,-8.5);
\path[->,ultra thick,blue] (-3.5,-7.9) edge [] (-4.5,-7.9);

\path (-4.5,-8.5) node[below,boxblue]{%
\begin{minipage}{0.39\textwidth}
Re-weighting ggh:
{\tt ggh@nnlo} top\\ contribution,
electro-weak contr.
\end{minipage}};


\path[->,ultra thick,blue] (2.5,-4.5) edge [bend left=40]
node[midway,right,black,boxblue]{
\begin{minipage}{0.22\textwidth}
\centering Diff. cross section
\end{minipage}
} (4.5,-6.5);

\path (4.5,-6.5) node[below,boxblue]{%
\begin{minipage}{0.39\textwidth}
\centering Calculation of ggh/bbh at \nblo{}
\end{minipage}};

\path[->,ultra thick,blue] (4.5,-7.25) edge [] (4.5,-8.5);

\path (4.5,-8.5) node[below,boxblue]{%
\begin{minipage}{0.39\textwidth}
\centering Performing cuts in $p_T,y,\eta$\\
ggh: $d\sigma/dp_T/dy/d\eta$
\end{minipage}};

\path (0,-10.) node[below,boxred]{%
\begin{minipage}{0.45\textwidth}
\centering Screen output/output file in SLHA-style
\end{minipage}};

\path[->,ultra thick,blue] (-4.5,-9.75) edge [bend right=40] (-3.9,-10.5);
\path[->,ultra thick,blue] (4.5,-9.75) edge [bend left=40] (3.9,-10.5);

\end{tikzpicture}
\caption{Internal workflow of \sushi{}.  Red boxes indicate interaction
  with the user, who has to provide an input and gets an output file, if
  no error messages are shown.  Green boxes refer to external code (see
  text), which is linked to/included in \sushi.}
\label{fig:workflow}
\end{figure}
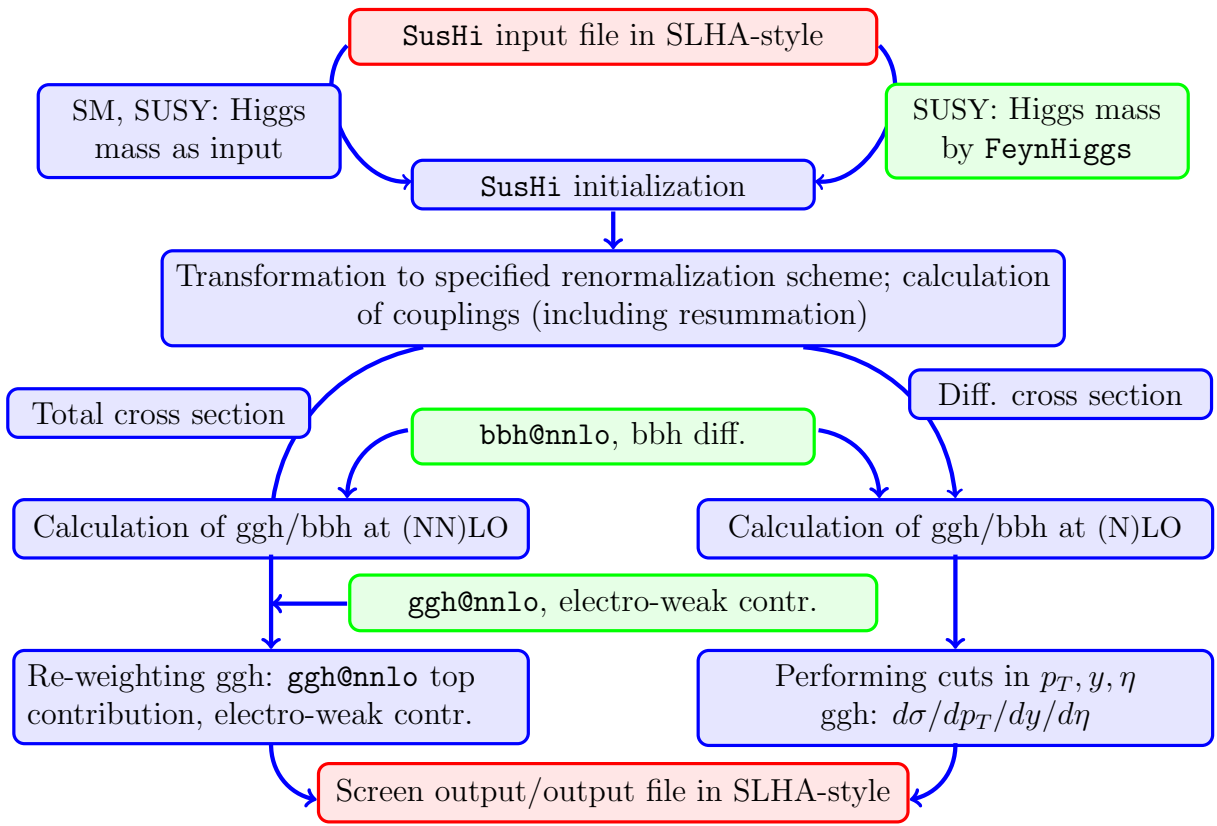

\subsection{External code}

As already mentioned, \sushi{} includes existing code like
{\tt ggh@nnlo}, {\tt bbh@nnlo}, or the electro-weak grid. The integration of
these programs does not require any action from the user though; they
are part of the distribution and are simply linked to \sushi{} upon
compilation.

However, \sushi{} can/must be linked to the following external code:
\begin{itemize}
\item {\tt FeynHiggs}
  \cite{Heinemeyer:1998np,Heinemeyer:1998yj,Degrassi:2002fi,Frank:2006yh}:
  For the calculation of the supersymmetric Higgs masses $\phi\in
  \lbrace h,H,A\rbrace$, \sushi{} can be linked to {\tt FeynHiggs}. Its
  input is controlled via the \sushi{} input files.  Note that the
  current version of \sushi{} does not support complex \mssm{}
  parameters, so {\tt FeynHiggs} is called with the default flags for
  the real \mssm{}.
\item {\tt LHAPDF} \cite{Whalley:2005nh}: \sushi{} has to be linked to
  {\tt LHAPDF} which provides a large variety of different \pdf{}
  sets. This allows one to change the \pdf{} set used by \sushi{} simply
  by changing the input file.
\end{itemize}


\subsection{Installation and Usage}
\label{sec:installation}

A tarball with the source files of \sushi{} can be obtained from
\citere{sushiaddress}. Unpacking results in a main folder with the
following subfolders:
\begin{description}
 \item[\tt bin]: contains the executable program {\tt sushi} after
   compilation
 \item[\tt example]: various example input files to be used in {\tt
   bin}
\item[\tt include, lib]: object files and libraries
\item[\tt src]: \sushi{} source files, including external code
\end{description}
The file {\tt README} in the main directory contains installation
instructions and the history of the code. Compilation is most easily
done by adjusting and running the Makefile in the main folder as
follows:
\begin{itemize}
\item Specify the location and the name of the library of {\tt LHAPDF},
  for example:\\ {\tt PDFLIBP = /usr/local/lib}\\ {\tt PDFLIB =
    -lLHAPDF}
\item If {\tt FeynHiggs} should be used, specify the main directory of
  the compiled {\tt FeynHiggs} library, for example:\\ {\tt FHPATH =
    /home/.../FeynHiggs-x.x.x}\\ Note that \sushi{} requires {\tt
    FeynHiggs} version 2.9 or higher.
\item Run the configure script in the main folder:\\
\begin{tikzpicture}[]
\node[below,right,boxblue]{%
\begin{minipage}{0.50\textwidth}
{\tt ./configure}
\end{minipage}
};\end{tikzpicture}\\ It tries to find the {\tt gfortran} or {\tt ifort}
compiler and the dependences on your local machine. If you prefer a
different compiler, or if the script fails, you can specify the relevant
variables ({\tt F77} and {\tt LDFLAGS}) in the file {\tt compilerissues}
yourself.
\item In the main folder, run make:\\
\begin{tikzpicture}[]
\node[below,right,boxblue]{%
\begin{minipage}{0.50\textwidth}
{\tt make [option]}
\end{minipage}
};\end{tikzpicture}\\ This command takes either of two optional
arguments: {\tt make predef=NoFeynHiggs} omits the link to {\tt
  FeynHiggs} which cannot be used in this case; {\tt make clean} deletes
all object files, libraries, and the executable {\tt sushi}.
\end{itemize}
After compilation, one may perform a test run of \sushi{} by copying one of
the input files from the {\tt example}- to the {\tt bin}-folder, change
to the {\tt bin}-folder, and run\\
\begin{tikzpicture}[]
\node[below,right,boxblue]{%
\begin{minipage}{0.50\textwidth}
   {\tt ./sushi sushi.in sushi.out}
\end{minipage}
};\end{tikzpicture}\\ Note that apart from the input filename, the user
also has to provide a name for the output file.  For parameter scans, we
recommend the auxiliary routines {\tt SLHAroutines}, which can be
downloaded from \citere{slharoutinesaddress}.  In the following we will
discuss the input and output files in more detail.


\subsection{Input and output files}\label{sec:inout}
\sushi{} allows for calculations in the \sm{} as well as in the
\mssm{}. Although the former is in many respects a limiting case of the
latter, \sushi{} distinguishes both cases, and we will discuss them
separately in what follows.

\subsubsection{Standard Model}
A typical input file for a \sm{} calculation is shown in the
following. It divides into blocks, each of which contains a number of
entries, specified by one or more leading blanks, an integer ``entry
number'', a value, and a comment initiated by the hash symbol \verb/#/.

\lstset{basicstyle=\scriptsize, frame=shadowbox}
{\tt
\begin{lstlisting}
Block SUSHI
  1   0		# model: 0 = SM, 1 = MSSM
  2   0		# 0 = scalar Higgs (h), 1 = pseudoscalar Higgs (A)
  3   0		# collider: 0 = p-p, 1 = p-pbar
  4   8000.d0	# center-of-mass energy in GeV
  5   2		# order ggh: -1 = off, 0 = LO, 1 = NLO, 2 = NNLO
  6   2 	# order bbh: -1 = off, 0 = LO, 1 = NLO, 2 = NNLO
  7   1 	# electroweak cont. for ggh:
                # 0 = no, 1 = light quarks at NLO, 2 = SM EW factor
Block SMINPUTS		# Standard Model inputs
  1   1.27934000e+02	# alpha_em^(-1)(MZ) SM MSbar
  2   1.16637000e-05	# G_Fermi
  3   1.17200000e-01	# alpha_s(MZ) SM MSbar
  4   9.11876000e+01	# m_Z(pole)
  5   4.20000000e+00	# m_b(m_b)
  6   1.73300000e+02	# m_t(pole)
  8   1.27500000e+00	# m_c(m_c)
Block MASS
  1   125.000000d0	# Higgs mass
Block DISTRIB
  1   0		# distribution : 0 = sigma_total, 1 = dsigma/dpt,
		#                2 = dsigma/dy,   3 = d^2sigma/dy/dpt
		#                (values for pt and y: 22 and 32)
  2   0		# pt-cut: 0 = no, 1 = pt > ptmin, 2 = pt < ptmax,
		#         3 = ptmin < pt < ptmax
 21   30.d0	# minimal pt-value ptmin in GeV
 22   100.d0	# maximal pt-value ptmax in GeV
  3   0		# rapidity-cut: 0 = no, 1 = Abs[y] < ymax,
		#    2 = Abs[y] > ymin, 3 = ymin < Abs[y] < ymax
 31   0.5d0	# minimal rapidity ymin
 32   1.5d0	# maximal rapidity ymax
  4   0		# 0 = rapidity, 1 = pseudorapidity
Block SCALES
  1   1.d0 	# renormalization scale muR/mh
  2   1.d0	# factorization scale muF/mh
Block RENORMBOT # Renormalization of the bottom sector
  1   0  	# m_b used for bottom Yukawa:     0 = OS, 1 = MSbar(mb), 2 = MSbar(muR)
Block PDFSPEC
  1   MSTW2008lo68cl.LHgrid		# name of pdf (lo)
  2   MSTW2008nlo68cl.LHgrid		# name of pdf (nlo)
  3   MSTW2008nnlo_asmzrange.LHgrid	# name of pdf (nnlo)
  4   0		# set number
Block VEGAS
  1   10000	# number of points
  2   5		# number of iterations
  3   10	# print: 0 = no output, 1 = prettyprint, 10 = table
Block FACTORS
  1   0.d0	# factor for yukawa-couplings: c
  2   1.d0	# t
  3   1.d0	# b
\end{lstlisting} }

{\tt Block SUSHI} specifies the crucial input for \sushi{}, namely the
model, the kind of Higgs boson to be considered (scalar or
pseudo-scalar\footnote{Apart from the scalar Higgs boson of the actual
  \sm{}, \sushi{} also provides results for a pseudo-scalar Higgs-like
  particle whose coupling to fermions is obtained from the corresponding
  \mssm{} couplings by setting $\tan\beta=1$.}), the type of collider,
the center-of-mass energy, the perturbative order for gluon fusion and
bottom-quark annihilation, and to which extent electro-weak corrections
to gluon fusion should be taken into account.

{\tt Block \sm{}INPUTS} contains the relevant \sm{} input.  We use
the electro-magnetic coupling $\alpha_{\text{EM}}$, Fermi's constant
$G_F$, and the $Z$-boson mass $m_Z$ (entries 1,2,4) to calculate the $W$
mass $m_W$ and the weak mixing angle $\sin\theta_W$. The input value for
$\alpha_s(m_Z)$ given in entry~3 is used for renormalization-group
({\abbrev RG}) running and {\abbrev RG} transformations.  We allow this
value to be different from the one required by the \pdf{}s which are
specified further below. The latter is taken from {\tt LHAPDF} and
enters the calculation as the coupling parameter of the perturbative
expansion of the cross section, for example \eqn{eq:crosssection} and
\neqn{eq:sig0}. The charm- and bottom-quark masses (entries 8,5) are to
be given in the $\msbar{}$ scheme as $m_c(m_c)$ and $\mbottom(\mbottom)$, while
the top-quark mass (entry 6) is required in the on-shell scheme. In the
\sm{}, the Higgs mass is a free parameter and has to be provided in
{\tt Block MASS}, entry 1.

{\tt Block DISTRIB} controls cuts or distributions with respect to the
transverse momentum $p_T$, the (pseudo-)rapidity $y$ ($\eta$), if
desired.  Note that differential cross sections (entry\,1 $\in\{
1,2,3\}$) can only be obtained for gluon fusion; in this case,
entries~22 and/or 32 specify the value of $p_T$ and/or $y$ ($\eta$).
Cuts are possible both for gluon fusion and bottom-quark annihilation;
they are applied by setting entry\,1 to 0 (``total cross section''), and
specifying entries\,2,21,22 and/or 3,31,32.  Note that entry\,4 changes
between rapidity $y$ and pseudo-rapidity $\eta$.

{\tt Block SCALES} defines the renormalization and factorization scales
relative to the Higgs mass. In accordance with \sct{sec:bottommasses},
{\tt Block RENORMBOT} offers different options for the renormalization
of the bottom Yukawa coupling; three options are currently implemented:
$\mbottom^Y\in\{\mbottom^\OS,\mbottom^{\msbar}(\mbottom),\mbottom^{\msbar}(\muR)\}$. {\tt
  Block PDFSPEC} contains the \pdf{} sets in the notation of {\tt
  LHAPDF}, consisting of the name of the \pdf{} grid file, and the set
number.  {\tt Block VEGAS} specifies integration parameters; note that
distributions or cuts require higher numerical precision than the total
cross section in order to reach comparable accuracy in the final result.
Finally, {\tt Block FACTORS} allows for additional factors in the Yukawa
couplings of the fermions. We add that also charm-quark contributions
can be taken into account by setting the corresponding factor to
$1$. Then the $c$-quark contributions at $\nblo{}$ are added using the
on-shell value $m_c^{\text{OS}}$ calculated from $\mcharm(\mcharm)$ as
done for the on-shell bottom-quark mass.  In case of the \mssm{}, for
which a detailed prescription follows, the charm-quark contributions can
be added as well.

\newpage
\subsubsection{Minimal Supersymmetric Standard Model}
In case of the \mssm{}, the input file contains a number of additional
{\tt Block}s. We show them here, together with the relevant modifications of
the \sm{} version:
{\tt 
\begin{lstlisting}
Block SUSHI
  1   1		# model: 0 = SM, 1 = MSSM
  2   1         # 0 = light Higgs (h), 1 = pseudoscalar (A), 2 = heavy Higgs (H)
[.....]
  5   2		# order ggh: -1 = off, 0 = LO, 1 = NLO, 2 = NNLO, 3 = ~NNLO stop for h
Block MINPAR    # SUSY breaking input parameters
  3   5.d0 	# tanb
Block EXTPAR
  3   800.d0  	# M_3
 11   2006.66d0 # A_t
 12   2006.66d0 # A_b
 23   200.d0 	# mu in GeV
 26   300.d0 	# M_A0
 43   1000.d0 	# M_Q3
 46   1000.d0 	# M_TR
 49   1000.d0 	# M_BR
Block FEYNHIGGS # FeynHiggs specific input
  1   0.d0 	# M_1
  2   200.d0 	# M_2
  3   2006.66d0	# A in GeV (except for A_t, A_b)
  4   1000.d0 	# M_SUSY in GeV (except for M_Q3, M_TR, M_BR)
Block RENORMBOT # Renormalization of the bottom sector
  1   0 	# m_b used for bottom Yukawa:  0 = OS, 1 = MSbar(m_b), 2 = MSbar(muR)
  2   1 	# tan(beta)-res. of Y_b: 0 = no, 1 = naive, 2 = full (for OS only)
  3   1 	# Delta_b: Take Delta_b from FeynHiggs: 0 = no, 1 = yes
Block RENORMSBOT # Renormalization of the sbottom sector
  1   2 	# m_b:     0 = OS, 1 = DRbar, 2 = dep; recommended: 2
  2   0 	# A_b:     0 = OS, 1 = DRbar, 2 = dep; recommended: 0
  3   0 	# theta_b: 0 = OS, 1 = DRbar         ; recommended: 0
Block FACTORS
  1   0.d0 # factor for yukawa-couplings: c
  2   1.d0 # t
  3   1.d0 # b
  4   1.d0 # st
  5   1.d0 # sb
\end{lstlisting} }

Entry\,2 of {\tt Block SUSHI} now distinguishes between the three
\mssm{} Higgs bosons. Entry\,5 allows to add approximated
\nnlo{} stop contributions for the light Higgs $h$.
{\tt Block MINPAR}, entry\,3 defines the value of
$\tan\beta$.  {\tt Block EXTPAR} fixes the parameters of the third
family of quarks and squarks in the \mssm{}. If the {\tt Block
FEYNHIGGS} is present, \sushi{} has to be linked to {\tt FeynHiggs}
(see \sct{sec:installation}) which will then be used to calculate the
Higgs masses from the parameters of that {\tt Block}.  In addition to
the \sm{} version, {\tt Block RENORMBOT} provides various ways of
resumming $t_\beta$-enhanced effects for the on-shell bottom Yukawa
coupling, see \sct{sec:bottommasses}. For a running coupling (entry\,1
$\in\{1,2\}$), resummation of those effects is always performed as shown
in \eqn{eq:msbarcoupling}.  {\tt Block RENORMSBOT} provides the choice
between the various options of \tab{tab:sbottomrenorm}.
The alternative
to {\tt Block FEYNHIGGS} is the specification of the Higgs masses and
the Higgs mixing angle $\alpha$ by hand; for example:
{\tt 
\begin{lstlisting}
Block ALPHA
 -2.58961078E-01	# mixing in Higgs sector
Block MASS
 25   125.216431E+00	# Higgs mass h
 26   303.288802E+00 	# Higgs mass H
 36   130.000000E+00	# Pseudoscalar Higgs mass A
\end{lstlisting}}
In this case, it is the user's responsibility to assure
consistency of the Higgs mass and the other parameters.
However, this option allows one to use
{\tt H3m}~\cite{Kant:2010tf}, for example, in order to take
into account three-loop effects to the \susy{} Higgs
mass~\cite{Harlander:2008ju,Kant:2010tf,Martin:2007pg}.

Three example input files can be found in the subfolder {\tt example},
namely a \sm{} input file and two \mssm{} input files, the latter two
for the usage with and without {\tt FeynHiggs}.

\subsubsection{Output file}

\sushi{} outputs the results of the calculation as well as some key
parameters derived from the input in the same format as the input
file. A typical example is shown here: {\tt
\begin{lstlisting}
Block SUSHIggh # Bon appetit
         1     1.65830627E+01   # ggh XS in pb   
Block SUSHIbbh # Bon appetit
         1     3.83211336E-01   # bbh XS in pb   
Block XSGGH # ggh MSSM-Cross sec. in pb
         2     1.21882475E+01   # NLO
[.....individual channels]
Block XSGGHEFF # ggh MSSM-Cross sec.
         1     1.44925736E+01   # ggh@NLO SM
         2     1.82888376E+01   # ggh@NNLO SM
         3     5.78917317E-02   # electroweak factor
Block XSBBH # bbh MSSM-Cross sec. in pb
         1     5.30129833E-01   # LO
         2     4.79035691E-01   # NLO
         3     3.83211336E-01   # NNLO
Block HGGSUSY # couplings of light Higgs h to 3. generation
[.....]
Block MASSOUT
         5     4.20000000E+00   # m_b(m_b), MSbar
        25     1.25216431E+02   # MSSM-Mh in GeV
[.....SM masses/sbottom/stop masses]
Block ALPHA # Effective Higgs mixing parameter
          -2.58961078E-01   # alpha
Block STOPMIX # stop mixing matrix
  1  1     7.07918788E-01   # V_11
  1  2    -7.06293841E-01   # V_12
  2  1     7.06293841E-01   # V_21
  2  2     7.07918788E-01   # V_22
[.....]
Block AD
  3  3     2.00666000E+03   # used Ab in GeV - def. accord. to scheme
Block AU
  3  3     2.00666000E+03   # used At in GeV
\end{lstlisting}}

The main result for the gluon fusion cross section, containing all
corrections specified by the user in the input file, is given as
entry\,1 in {\tt Block SUSHIggh}, the one for bottom-quark annihilation
in {\tt Block SUSHIbbh}. Individual contributions to the cross sections
are listed in {\tt Block XSGGH} and {\tt Block XSBBH}; their meaning
should be obvious from the comment in the output file. Note
that the results denoted ``{\tt LO}'' etc.\ mean that the \lo{} partonic
cross section is convolved with the \pdf{} set given in entry\,1 of {\tt
Block PDFSPEC} in the input file.

For gluon fusion, the {\tt Block XSGGHEFF} contains the \nnlo{}
top-(s)quark results as obtained by {\tt ggh@nnlo}, and the electro-weak
correction factor as determined in \sct{sec:ewcontributions}.

In addition, {\tt Block HGGSUSY} lists the non-resummed \mssm{}
couplings of the quarks and squarks to the Higgs boson under
consideration. {\tt Block MASSOUT} gives the relevant \sm{} and \susy{}
masses as well as the Higgs mass. Not shown above is
the {\tt Block INTERNALMASSES}, which provides the different bottom
masses entering the calculation of gluon-fusion cross sections, and
{\tt SCALESOUT} showing the renormalization and factorization scale as well
as the value of $\alpha_s(\muR)$ taken from the \pdf{} set at \nblo{}.
Finally all output files have the corresponding input file attached at
the end.



\section{Conclusion}
\label{sec:conclusions}

In this article we described the {\tt Fortran} code \sushi{} for the
calculation of the cross section for Higgs production in gluon fusion
and bottom-quark annihilation at hadron colliders. It works both in the
\sm{} and the \mssm{}, evaluates inclusive cross sections,
distributions, and allows for kinematical cuts on the Higgs
4-momentum. It includes higher order \qcd{} and electro-weak corrections
and takes into account the effect from squarks and gluinos.

\sushi{} allows one to choose among various renormalization schemes for
the sbottom sector and the bottom Yukawa coupling, and includes the
resummation of $\tan\beta$-enhanced effects.  For the calculation of the
Higgs mass in the \mssm{} \sushi{} can be linked to {\tt FeynHiggs}.
\sushi{} can be downloaded from \citere{sushiaddress}.


\paragraph{Acknowledgments}

This work was supported by DFG, contract C0214101A and the Helmholtz
Alliance ``Physics at the Terascale''.

We are grateful to the authors of
\citeres{Degrassi:2008zj,Degrassi:2010eu,Degrassi:2011vq,Degrassi:2012vt,Bagnaschi:2011tu},
in particular Pietro Slavich for many helpful comments on our program,
this manual, for many cross checks, and for fruitful discussions.
Regarding the interpretation of {\tt FeynHiggs} internal and output
variables our thanks go to Thomas Hahn, Sven Heinemeyer, Heidi Rzehak
and Georg Weiglein.  Concerning the electro-weak contributions of light
quarks, we thank Alessandro Vicini and Giuseppe Degrassi for comparing
numbers.

\newpage


\begin{appendix}

\section{Formulas: Higgs-squark couplings}
\label{sec:higgssquark}

In this section we present the couplings of the three neutral Higgs
bosons $\phi$ of the \mssm{} to the quarks and squarks being implemented in
\sushi{}. The relevant Feynman rules can be written in the form
\begin{align}
 \parbox{27mm}{\includegraphics[width=0.2\textwidth]{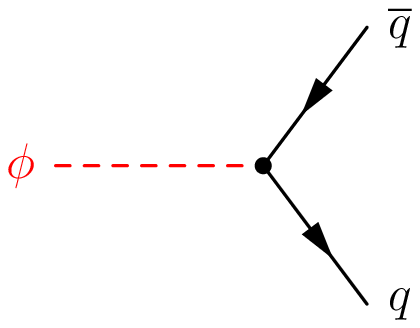}}=i\frac{\mass{q}}{v}g_q^\phi\qquad\text{and}\qquad
 \parbox{27mm}{\includegraphics[width=0.2\textwidth]{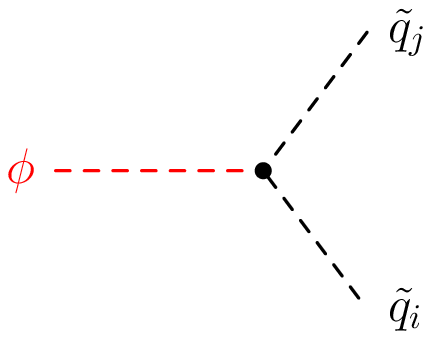}}=i\frac{\mass{q}^2}{v}g_{\tq,ij}^\phi\quad,
\end{align}
where $v=2m_W/g=1/\sqrt{\sqrt{2}G_F}=\sqrt{v_d^2+v_u^2}$. The couplings $g_q^\phi$ of the
Higgs boson $\phi$ to the quarks $q$ with respect to the \sm{} Higgs boson coupling
were already presented in \eqn{eq:couplfermion}. The couplings $g_{\tq,ij}^\phi$
of the squarks to the light and heavy Higgs can be split in the form
\begin{align}
 g_{\tq,ij}^\phi=g_{\tq,ij}^{\phi,\EW}+g_{\tq,ij}^{\phi,\mu}+g_{\tq,ij}^{\phi,\alpha}\quad.
\end{align}
In case of the light Higgs $h$ we obtain for the couplings:
\begin{align}
 &g_{\ttop,11}^{h,\EW}=c_{\ttop,1}^{\EW}c^2_{\theta_\ttop}+c_{\ttop,2}^{\EW}s^2_{\theta_\ttop} &
 &g_{\tbot,11}^{h,\EW}=c_{\tbot,1}^{\EW}c^2_{\theta_\tbot}+c_{\tbot,2}^{\EW}s^2_{\theta_\tbot} 
 \\
 &g_{\ttop,22}^{h,\EW}=c_{\ttop,1}^{\EW}s^2_{\theta_\ttop}+c_{\ttop,2}^{\EW}c^2_{\theta_\ttop}&
 &g_{\tbot,22}^{h,\EW}=c_{\tbot,1}^{\EW}s^2_{\theta_\tbot}+c_{\tbot,2}^{\EW}c^2_{\theta_\tbot}
 \\
 &g_{\ttop,12}^{h,\EW}=g_{\ttop,21}^{h,\EW}=\frac{1}{2}\left(c_2^{\EW}-c_1^{\EW}\right)s_{2\theta_\ttop}&
 &g_{\tbot,12}^{h,\EW}=g_{\tbot,21}^{h,\EW}=\frac{1}{2}\left(c_2^{\EW}-c_1^{\EW}\right)s_{2\theta_\tbot}
 \\
 &g_{\ttop,11}^{h,\mu}=-g_{\ttop,22}^{h,\mu}=\frac{\mu}{\mass{t}}\frac{\cos(\alpha-\beta)}{s^2_\beta}s_{2\theta_\ttop}&
 &g_{\tbot,11}^{h,\mu}=-g_{\tbot,22}^{h,\mu}=-\frac{\mu}{\mbottom}\frac{\cos(\alpha-\beta)}{c^2_\beta}s_{2\theta_\tbot} 
 \\
 &g_{\ttop,12}^{h,\mu}=g_{\ttop,21}^{h,\mu}=\frac{\mu}{\mass{t}}\frac{\cos(\alpha-\beta)}{s^2_\beta}c_{2\theta_\ttop}&
 &g_{\tbot,12}^{h,\mu}=g_{\tbot,21}^{h,\mu}=-\frac{\mu}{\mbottom}\frac{\cos(\alpha-\beta)}{c^2_\beta}c_{2\theta_\tbot}
 \\
 &g_{\ttop,11}^{h,\mu}=\frac{c_\alpha}{s_\beta}\left(2+\frac{\mstop{1}^2-\mstop{2}^2}{2\mass{t}^2}s^2_{2\theta_\ttop}\right)&
 &g_{\tbot,11}^{h,\mu}=-\frac{s_\alpha}{c_\beta}\left(2+\frac{\msbottom{1}^2-\msbottom{2}^2}{2\mbottom^2}s^2_{2\theta_\tbot}\right) 
 \\
 &g_{\ttop,22}^{h,\mu}=\frac{c_\alpha}{s_\beta}\left(2-\frac{\mstop{1}^2-\mstop{2}^2}{2\mass{t}^2}s^2_{2\theta_\ttop}\right)&
 &g_{\tbot,22}^{h,\mu}=-\frac{s_\alpha}{c_\beta}\left(2-\frac{\msbottom{1}^2-\msbottom{2}^2}{2\mbottom^2}s^2_{2\theta_\tbot}\right) 
 \\
 &g_{\ttop,12}^{h,\mu}=g_{\ttop,21}^{h,\mu}=
 \frac{c_\alpha}{s_\beta}\frac{\mstop{1}^2-\mstop{2}^2}{2\mass{t}^2}s_{2\theta_\ttop}c_{2\theta_\ttop} &
 &g_{\tbot,12}^{h,\mu}=g_{\tbot,21}^{h,\mu}=
 -\frac{s_\alpha}{c_\beta}\frac{\msbottom{1}^2-\msbottom{2}^2}{2\mbottom^2}s_{2\theta_\tbot}c_{2\theta_\tbot}
\end{align}
Therein we made use of the abbreviations $s_x=\sin x$ and $c_x=\cos x$ and defined:
\begin{align}
 &c_{\ttop,1}^{\EW} = -\frac{m_Z^2}{\mass{t}^2}\left(1-\frac{4}{3}s^2_{\theta_W}\right)\sin(\alpha+\beta)&
 &c_{\tbot,1}^{\EW} = \frac{m_Z^2}{\mbottom^2}\left(1-\frac{2}{3}s^2_{\theta_W}\right)\sin(\alpha+\beta) 
 \\
 &c_{\ttop,2}^{\EW} = -\frac{m_Z^2}{\mass{t}^2}\frac{2}{3}s^2_{\theta_W}\sin(\alpha+\beta)&
 &c_{\tbot,2}^{\EW} = \frac{m_Z^2}{\mbottom^2}\frac{2}{3}s^2_{\theta_W}\sin(\alpha+\beta)
\end{align}
The couplings to the heavy Higgs $H$ are easy to obtain
by the replacement $\alpha\rightarrow \alpha - \tfrac{\pi}{2}$
in the previous formulas. In case of the {\abbrev CP}-odd Higgs $A$
the couplings are given by:
\begin{align}
 &g_{\ttop,11}^A=g_{\ttop,22}^A =g_{\tbot,11}^A = g_{\tbot,22}^A=0\\
 &g_{\ttop,12}^A=-g_{\ttop,21}^A=\frac{1}{t_\beta}\frac{\mstop{1}^2-\mstop{2}^2}{2\mass{t}^2}s_{2\theta_\ttop}
 +\frac{\mu}{\mass{t}}\left(1+\frac{1}{t_\beta^2}\right)\\
 &g_{\tbot,12}^A=-g_{\tbot,21}^A=t_\beta\frac{\msbottom{1}^2-\msbottom{2}^2}{2\mbottom^2}s_{2\theta_\tbot}
 +\frac{\mu}{\mbottom}\left(1+t_\beta^2\right) 
\end{align}
We add that $\mbottom$ is partially interpreted as the bottom mass in
the sbottom sector, namely where it is meant to be part of the
Higgs-sbottom coupling.

\end{appendix}


\def\app#1#2#3{{\it Act.~Phys.~Pol.~}\jref{\bf B #1}{#2}{#3}}
\def\apa#1#2#3{{\it Act.~Phys.~Austr.~}\jref{\bf#1}{#2}{#3}}
\def\annphys#1#2#3{{\it Ann.~Phys.~}\jref{\bf #1}{#2}{#3}}
\def\cmp#1#2#3{{\it Comm.~Math.~Phys.~}\jref{\bf #1}{#2}{#3}}
\def\cpc#1#2#3{{\it Comp.~Phys.~Commun.~}\jref{\bf #1}{#2}{#3}}
\def\epjc#1#2#3{{\it Eur.\ Phys.\ J.\ }\jref{\bf C #1}{#2}{#3}}
\def\fortp#1#2#3{{\it Fortschr.~Phys.~}\jref{\bf#1}{#2}{#3}}
\def\ijmpc#1#2#3{{\it Int.~J.~Mod.~Phys.~}\jref{\bf C #1}{#2}{#3}}
\def\ijmpa#1#2#3{{\it Int.~J.~Mod.~Phys.~}\jref{\bf A #1}{#2}{#3}}
\def\jcp#1#2#3{{\it J.~Comp.~Phys.~}\jref{\bf #1}{#2}{#3}}
\def\jetp#1#2#3{{\it JETP~Lett.~}\jref{\bf #1}{#2}{#3}}
\def\jphysg#1#2#3{{\small\it J.~Phys.~G~}\jref{\bf #1}{#2}{#3}}
\def\jhep#1#2#3{{\small\it JHEP~}\jref{\bf #1}{#2}{#3}}
\def\mpl#1#2#3{{\it Mod.~Phys.~Lett.~}\jref{\bf A #1}{#2}{#3}}
\def\nima#1#2#3{{\it Nucl.~Inst.~Meth.~}\jref{\bf A #1}{#2}{#3}}
\def\npb#1#2#3{{\it Nucl.~Phys.~}\jref{\bf B #1}{#2}{#3}}
\def\nca#1#2#3{{\it Nuovo~Cim.~}\jref{\bf #1A}{#2}{#3}}
\def\plb#1#2#3{{\it Phys.~Lett.~}\jref{\bf B #1}{#2}{#3}}
\def\prc#1#2#3{{\it Phys.~Reports }\jref{\bf #1}{#2}{#3}}
\def\prd#1#2#3{{\it Phys.~Rev.~}\jref{\bf D #1}{#2}{#3}}
\def\pR#1#2#3{{\it Phys.~Rev.~}\jref{\bf #1}{#2}{#3}}
\def\prl#1#2#3{{\it Phys.~Rev.~Lett.~}\jref{\bf #1}{#2}{#3}}
\def\pr#1#2#3{{\it Phys.~Reports }\jref{\bf #1}{#2}{#3}}
\def\ptp#1#2#3{{\it Prog.~Theor.~Phys.~}\jref{\bf #1}{#2}{#3}}
\def\ppnp#1#2#3{{\it Prog.~Part.~Nucl.~Phys.~}\jref{\bf #1}{#2}{#3}}
\def\rmp#1#2#3{{\it Rev.~Mod.~Phys.~}\jref{\bf #1}{#2}{#3}}
\def\sovnp#1#2#3{{\it Sov.~J.~Nucl.~Phys.~}\jref{\bf #1}{#2}{#3}}
\def\sovus#1#2#3{{\it Sov.~Phys.~Usp.~}\jref{\bf #1}{#2}{#3}}
\def\tmf#1#2#3{{\it Teor.~Mat.~Fiz.~}\jref{\bf #1}{#2}{#3}}
\def\tmp#1#2#3{{\it Theor.~Math.~Phys.~}\jref{\bf #1}{#2}{#3}}
\def\yadfiz#1#2#3{{\it Yad.~Fiz.~}\jref{\bf #1}{#2}{#3}}
\def\zpc#1#2#3{{\it Z.~Phys.~}\jref{\bf C #1}{#2}{#3}}
\def\ibid#1#2#3{{ibid.~}\jref{\bf #1}{#2}{#3}}
\def\otherjournal#1#2#3#4{{\it #1}\jref{\bf #2}{#3}{#4}}
\newcommand{\jref}[3]{{\bf #1}, #3 (#2)}
\newcommand{\hepph}[1]{[hep-ph/#1]}
\newcommand{\mathph}[1]{[math-ph/#1]}
\newcommand{\arxiv}[2]{[arXiv:#1]}
\newcommand{\bibentry}[4]{#1, ``#2'', #3\ifthenelse{\equal{#4}{}}{}{ }#4.}
\bibliographystyle{h-physrev}


\end{document}